\documentclass[journal]{IEEEtran}

\usepackage{lineno,hyperref}
\usepackage{siunitx}
\usepackage{amsmath}
\usepackage{amsmath,bm}
\usepackage{xcolor}
\usepackage{multirow}
\usepackage{comment}
\usepackage{subcaption}
\usepackage{makecell}
\usepackage{graphicx}
\usepackage{gensymb}

\newsavebox{\bigimage}

\begin{document}

\title{Investigation of Inward-Outward Ring Permanent Magnet Array for Portable Magnetic Resonance Imaging (MRI)}

\author{\IEEEauthorblockN{Ting-Ou LIANG\IEEEauthorrefmark{1},
MinXuan XU\IEEEauthorrefmark{2},
Wenwei YU\IEEEauthorrefmark{2}, and
Shao Ying HUANG\IEEEauthorrefmark{1,3}}
\IEEEauthorblockA{\IEEEauthorrefmark{1}Singapore University of Technology and Design,
8 Somapah Road, Singapore 487372}
\IEEEauthorblockA{\IEEEauthorrefmark{2}Center for Frontier Medical Engineering, Chiba University, Inage Ku, Yayoi Cho, 1-33, Chiba, 263-8522, Japan}
\IEEEauthorblockA{\IEEEauthorrefmark{3}National University of Singapore, IE Kent Ridge Road, Singapore 119228}

\thanks{Corresponding author: Shao Ying HUANG (email: huangshaoying@sutd.edu.sg).}}

\maketitle

\begin{abstract}
Permanent magnet array (PMA) is a popular option to provide the main magnetic field in a dedicated portable magnetic resonance imaging (MRI) system because it does not need power or a cooling system and has a much stronger field strength compared to a resistive magnet.
Aside from the popular Halbach array that has a transversal field direction, the Inward-Outward ring (IO ring) array is a promising candidate that offers a longitudinal field direction with various design and engineering possibilities. 
In this article, a thorough study of IO ring arrays is conducted by examining 
the relation between the design parameters and its field patterns, its variants that lead to different applications and their properties. 
A detailed comparison between an IO ring array and Halbach array was conducted and reported. 
Moreover, the feasibility of building an IO ring array in a lab is demonstrated. 
The investigations strongly indicate that IO ring is a promising candidate that can offer high and homogeneous fields or a desired field pattern to portable MRI systems. With a longitudinal field direction, an IO ring array opens up opportunities to adopt MRI advanced technology and techniques in a portable system to improve image quality and shorten scan time.
\end{abstract}


\section{Introduction}
\label{sec:intro}

Permanent magnet arrays (PMAs) are widely used for portable magnetic resonance imaging (MRI) due to low power consumption and small footprint\,\cite{cooley2018design,ren2015magnet,oreilly2019halbach}. Compared to a superconducting magnet, a PMA does not require a sophisticated cooling system for operation though the field strength is compromised. Compared to a resistive electromagnet, it has higher field strength but no heat dissipation. 

For MRI, a PMA can be used to supply a homogeneous field\,\cite{oreilly2019halbach}, a gradient field\,\cite{LIANG2022IORing}, or a combination of the above\,\cite{cooley2015,LIANG2022IORing}.
To generate a homogeneous field, \textit{in-situ} PMAs, PMA for imaging inside the array\,\cite{huang2019_iMRI}, can be used. 
A widely used type of magnet is the C-shaped\,\cite{Cheng2001_CShape}/H-shaped PMA, which comprises two poles (with aggregated magnets), pole faces, and an iron yoke. It offers dipolar magnetic fields between the two pole faces for imaging\,\cite{siemens}. The distance between the pole faces can go up to 80\,cm\,\cite{Srinivas2022dipolar}. 
For the recent effort on low-field portable MRI, a cylindrical magnet array is popular where imaging is done in the bore. Examples are a Halbach array\,\cite{halbach1980design} with dipolar transversal fields and an inward-outward (IO) ring array with dipolar longitudinal fields\,\cite{Nishino1983singleRing,Miyajima1985Ring_pair,aubert1994permanent,ren2018design}. 
The former is more well-known than the latter, but the latter offers unique features.
In the literature, comprehensive reviews on permanent magnets and PMAs for MRI are presented\,\cite{huang2019_iMRI,Blumer2016HalbachBook,Johns2016MobileNMRBook}.

The inward-outward (IO) ring array is firstly proposed by G. Miyajima back in 1985\,\cite{Miyajima1985Ring_pair}, which is also referred to as ``spokes-and-hub'' magnets in some literature\,\cite{kuang2018}. It consists of a ring pair with one ring having inward polarization, and the other one having outward polarization, where the inwardly polarized ring was first introduced by E. Nishino in 1983 for magnetic medical applications\,\cite{Nishino1983singleRing}. An IO ring array supplies concentric longitudinal magnetic field within the enclosed cylindrical region, which can achieve high field strength.

In the literature, different designs were proposed based on an
O ring. In\,\cite{aubert1991cylindrical}, the superposition of multiple IO ring arrays was proposed by G. Aubert to obtain a homogeneous field pattern within a bore with a diameter of 40\,cm. Ferromagnetic yoke was introduced to be placed over the IO ring array to confine the magnetic flux in the cylindrical bore\,\cite{Miyajima1985Ring_pair}. In recent years, the irregular-shaped IO ring was proposed, offering a field pattern with built-in monotonic gradient, which can be rotated for signal encoding for imaging\,\cite{ren2019irregular}. It has high average field, however, it consists of dense arch-shaped magnets that are hard to implement. Recently, an IO ring-based PMA for head imaging consisting of magnet cuboids was designed using physics-guided optimizations which shows high average field strength and linear field pattern for signal encoding with light weight\,\cite{LIANG2022IORing}.

IO ring array supplies magnetic field in the axial direction, thus it can work with high performance coils designed for the existing MRI systems\,\cite{tommy2012coil} and allow easy pre-polarization\,\cite{lurie2010paper}. It has decent space between the two rings for potential interventions. It can have design variance for various applications, e.g., single-sided array reported in \,\cite{liang2023ismrm} for spine imaging. It has much unrevealed potential that remains understudied. 
In this paper, investigations on the relation of the design parameters of IO ring and its field patterns are presented, and the potential of this type of magnet configuration is explored. Several variants of IO ring array 
that can lead to different applications and their properties are examined. A detailed comparison between an IO ring array and a Halbach array was conducted and presented. Moreover, the feasibility of building an IO ring array in a lab is introduced and demonstrated. 

\section{Properties of IO Ring}
\label{sec:properties}

\subsection{IO ring and its discretization}
\label{subsec:discretize}
\begin{figure}[t]
\centering
\begin{center}
\newcommand{\patchSize}{2.45cm}
\scriptsize
\setlength\tabcolsep{0.1cm}
\includegraphics[width=0.8\linewidth]{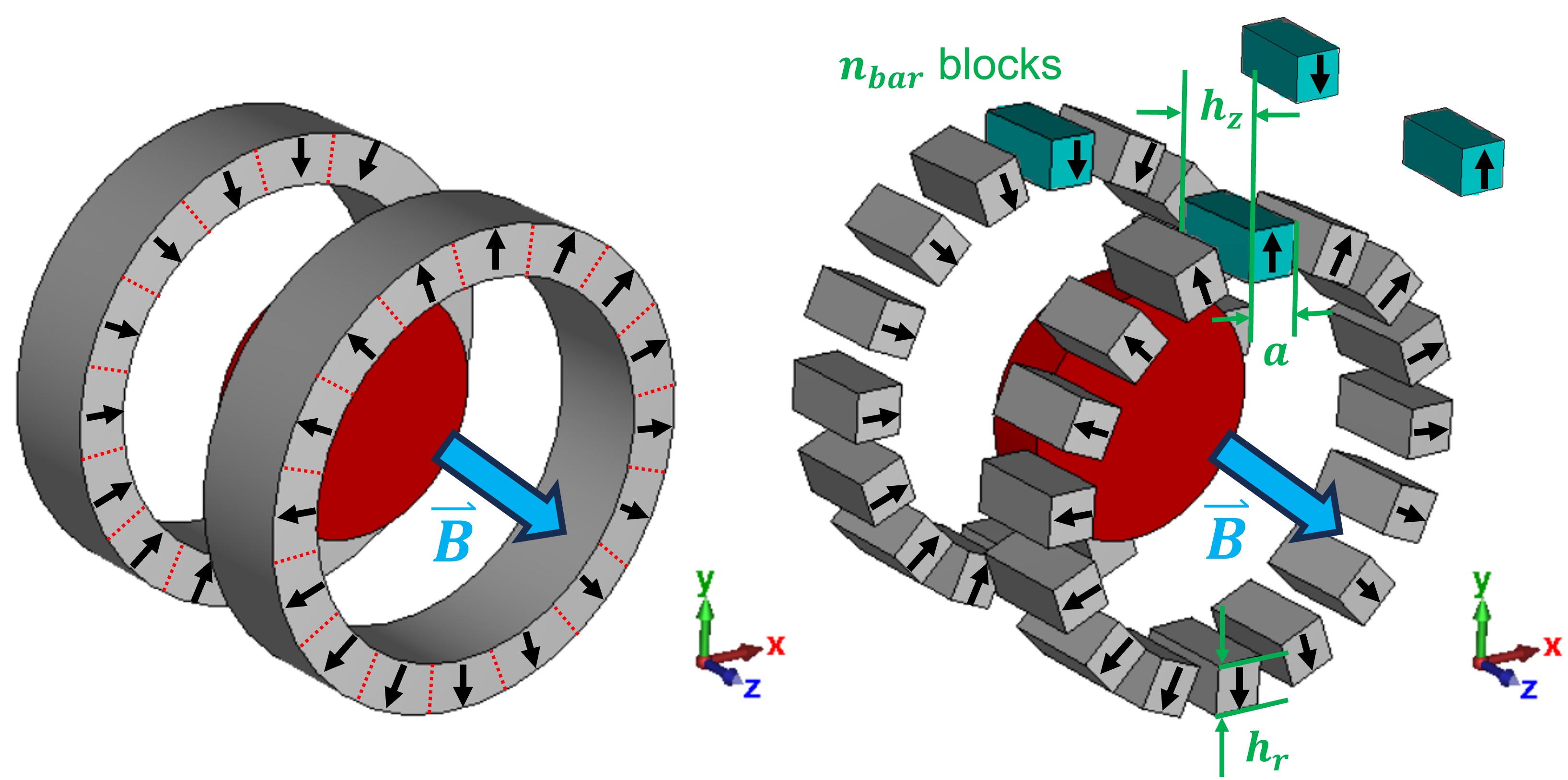}\\
(a)\hspace{0.45\linewidth}(b)\\    \includegraphics[width=0.8\linewidth]{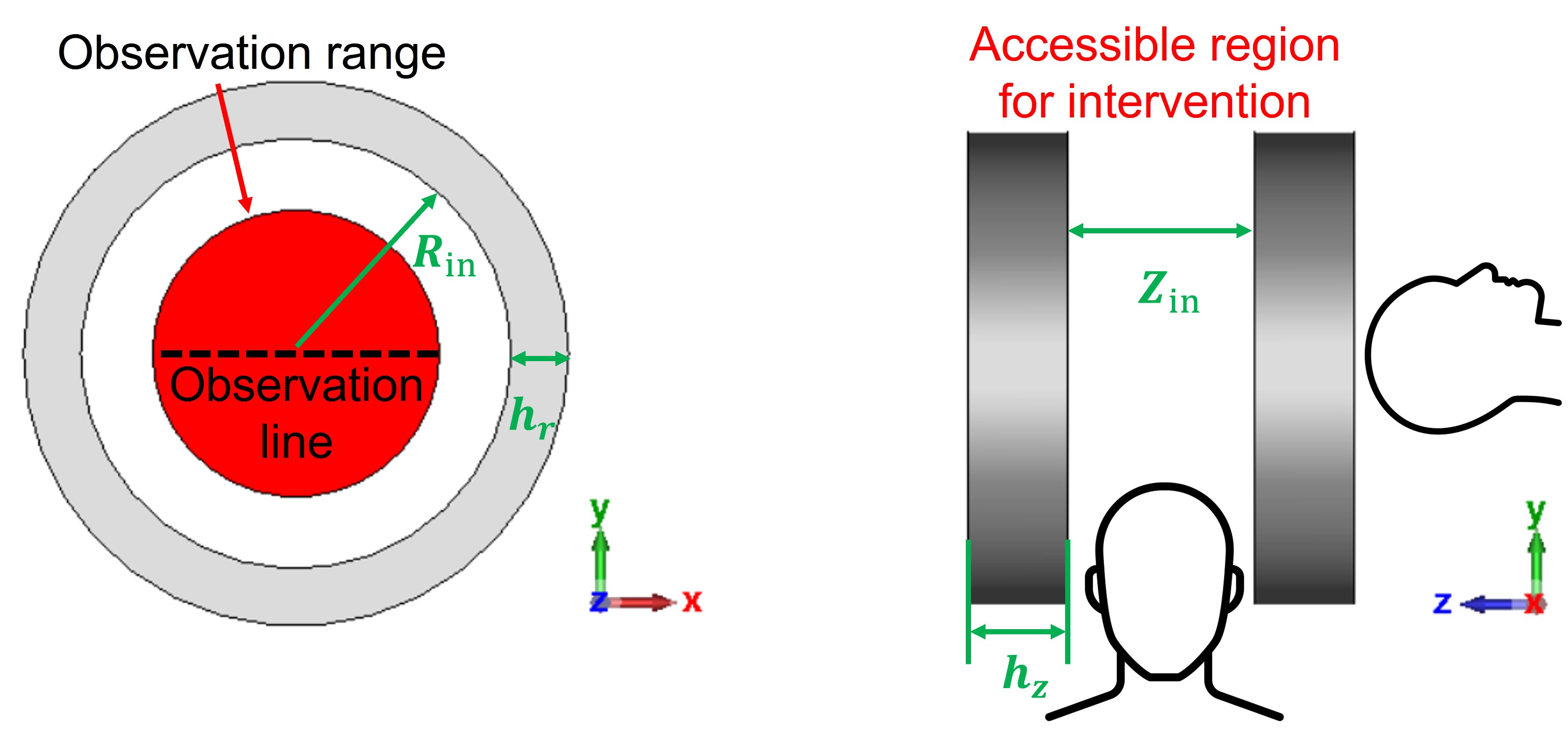}\\
(c)\hspace{0.45\linewidth}(d)\\ 	
\end{center}
\caption{A simple IO ring array composed of one outward-polarized ring and one inward-polarized ring, (a) 3D view, (b) the discretized IO ring array with cuboid magnets, (c) the cross-sectional view on the $xy$-plane, (d) the side view on the $yz$-plane.}
\label{fig:overview}
\end{figure}

Fig.\,\ref{fig:overview}\,(a), (c), and (d) show the 3D view and two side views of a basic IO ring array. 
The magnet can be accessed for scanning in the ways shown in Fig.\,\ref{fig:overview}\,(d). Meanwhile, as shown, the gap between the rings allows access for intervention. 
There are four parameters defining the geometry of the IO ring array, as shown in Fig.\,\ref{fig:overview}\,(c) and (d), the inner radius and the thickness of the ring, $R_\text{in}$ and $h_\text{r}$, respectively, the distance between the inner surfaces of the two rings, $Z_\text{in}$, and the thickness of each ring, $h_\text{z}$. Theoretically, an IO ring array can be implemented by arc-shaped magnets indicated by Fig.\,\ref{fig:overview}\,(a). However, in practice, arc-shaped magnets are less available. Therefore, the IO ring is usually discretized using cuboid magnets\,\cite{aubert1991cylindrical}, as shown in Fig.\,\ref{fig:overview}\,(b) with $n_\text{bar}$ blocks for each ring. The circumferential edge length, width, and thickness of a cuboid are denoted using $a$, $h_r$, and $h_z$, respectively. 

\begin{figure}[t]
\centering
\begin{center}
\newcommand{\patchSize}{2.45cm}
\scriptsize
\setlength\tabcolsep{0.1cm}
\includegraphics[width=\linewidth]{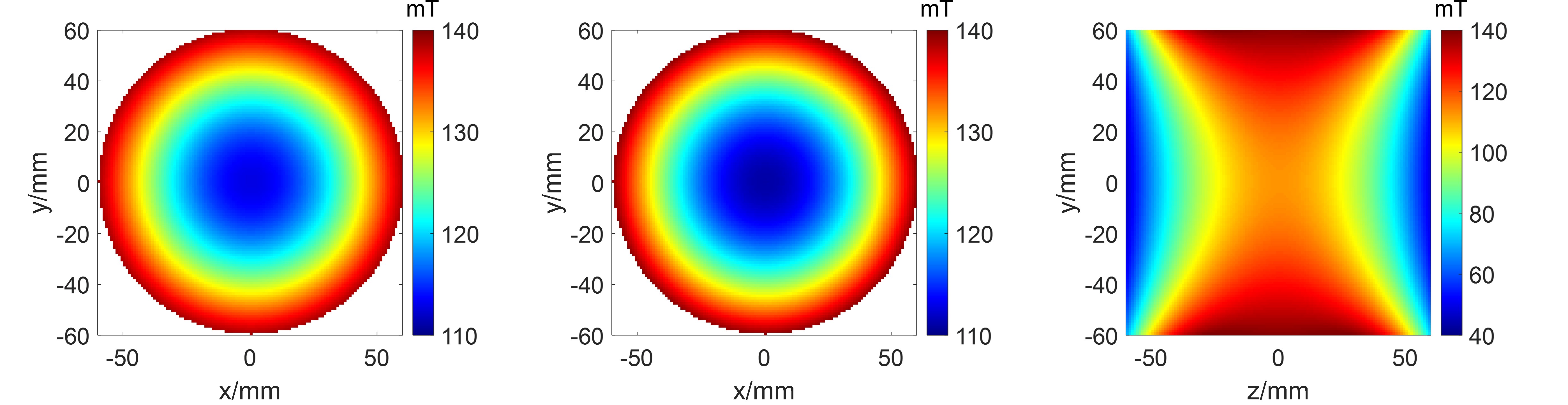}\\
(a)\hspace{0.3\linewidth}(b)\hspace{0.3\linewidth}(c)\\
\includegraphics[width=\linewidth]{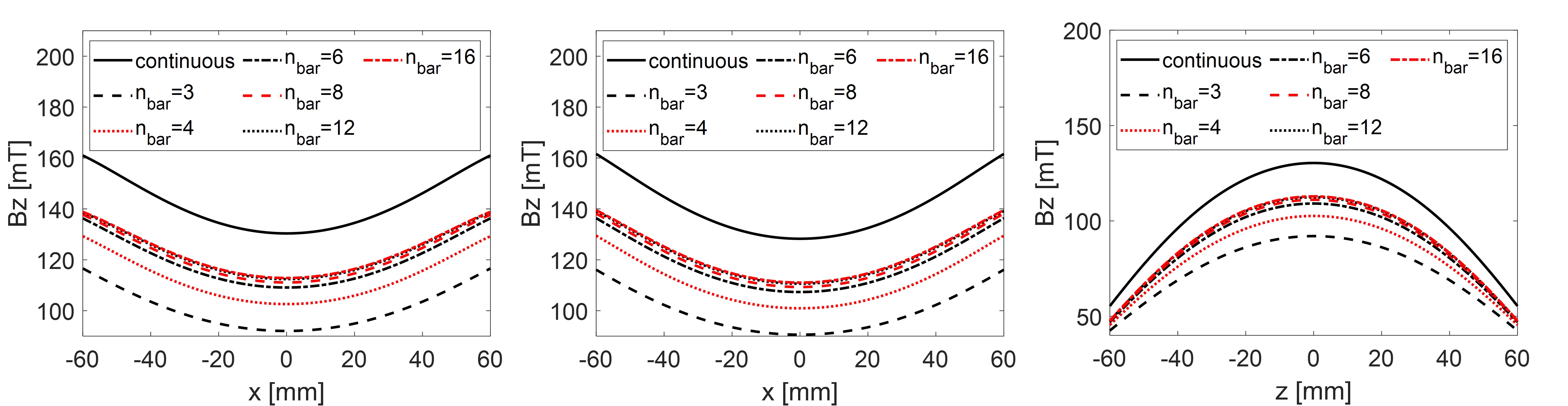}\\
(d)\hspace{0.3\linewidth}(e)\hspace{0.3\linewidth}(f)\\
\end{center}
\caption{The $B_\text{z}$ magnetic field component when $n_\text{bar}=16$ at (a) $z=0$, (b) $z=10$ mm, and (c) $x=0$. (d)-(f) are the magnetic field when sweeping through $n_\text{bar}=0,3,4,6,8,12,16$ along (d)-(e) the $x$-axis and (f) the $z$-axis.}
\label{fig:result_nbar}
\end{figure}

The effect of discretization is investigated. In this study, $n_\text{bar}$ was set to 3, 4, 6, 8, 12, and 16, $R_\text{in}=100$\,mm, $Z_\text{in}=100$\,mm, $h_\text{z}=40$\,mm, and $h_\text{r}=40$\,mm.
$n_\text{bar}$ affects the edge length $a$, $a=2R_\text{in}\tan(\pi/n_\text{bar})$ so that the magnets are placed head-to-tail with the maximum possible volume. The observation region, i.e., field-of-view (FoV), was set to be a cylinder with a radius of 60\,mm, a length of 20\,mm, and the axis aligned with that of the IO ring array. 
The magnetic fields generated by IO ring arrays at different levels of discretization are compared to that of a continuous IO ring array with the same dimensions. MagTetris\,\cite{LIANG2023MagTetris} was used for the calculation of the magnetic field. 

Fig.\,\ref{fig:result_nbar}\,(a) and (b) shows the 2D plot of the $z$-components of the magnetic fields at $n_\text{bar}=16$ on the $xy$-plane at z\,=\,0\,mm and 10\,mm, respective, and Fig.\,\ref{fig:result_nbar}\,(c) shows those on the $yz$-plane. As show, regardless the location along the $z$-axis, the field pattern on the $xy$-plane is concentric. Therefore, 1D field plots along the radial direction offers sufficient information on field strength and homogeneity. The field patterns of other cases with a different $n_\text{bar}$ are all concentric, except the case at $n_\text{bar}=3$ that the magnet blocks are the lest axial symmetric. Thus, the effect of $n_\text{bar}$ on field distributions is further examined by checking the 1D field distributions of the cases under investigation. 

Fig.\,\ref{fig:result_nbar}\,(a) - (b) show the calculated $z$-components of the magnetic fields along the $x$-axis (as shown in Fig.\,\ref{fig:result_nbar}\,(a)) and Fig.\,\ref{fig:result_nbar}\,(c) shows the $z$-component fields in the $z$-direction. 
In Fig.\,\ref{fig:result_nbar}\,(a) - (c), all the curves exhibit similar curvatures, the field strength increases as $n_\text{bar}$ increases and shows a clear convergence. 
Further examining the fields along the $x$-direction, as shown in Fig.\,\ref{fig:result_nbar}\,(a) - (b), all cases show similar gradients from the center to the periphery of the circular FoV (in the range of 103 - 113\,mT/m at $z$\,=\,0\,mm and the range of 126 - 138\,mT/m at $z$\,=\,10\,mm). This indicates a homogeneous field is hard to obtain by only changing $n_\text{bar}$. Comparing the cases at $z$\,=\,0\,mm in Fig.\,\ref{fig:result_nbar}\,(a) to those at $z$\,=\,10\,mm in Fig.\,\ref{fig:result_nbar}\,(b), besides the gradients, they are similar in the mean magnetic field when $n_\text{bar}$ is the same.
For the fields along the $z$-direction, there is monotonic gradients between $z$\,=\,0\,mm to $\pm$60\,mm which increases when $n_\text{bar}$ increases. 

Comparing the curves at $n_\text{bar}=16$ to those of the continuous case, it is noticed that there is still a field difference of 2\,mT between the field at $n_\text{bar}=16$ and that of the continuous case. 
The difference is caused by the limited filling factor of the discretized cases, a filling factor of less than one if their continuous counterpart is considered to have a filling factor of one. It implies that when magnet cuboids are used to implement an IO ring array, it is inevitable that the field strength is compromised, which can be considered as the price paid for availability of building materials. 


\subsection{Effect of other design parameters on magnetic field pattern}
\label{subsec:param_study}

\begin{figure}[thp]
\centering
\begin{center}
\newcommand{\patchSize}{2.45cm}
\scriptsize
\setlength\tabcolsep{0.1cm}
\includegraphics[width=\linewidth]{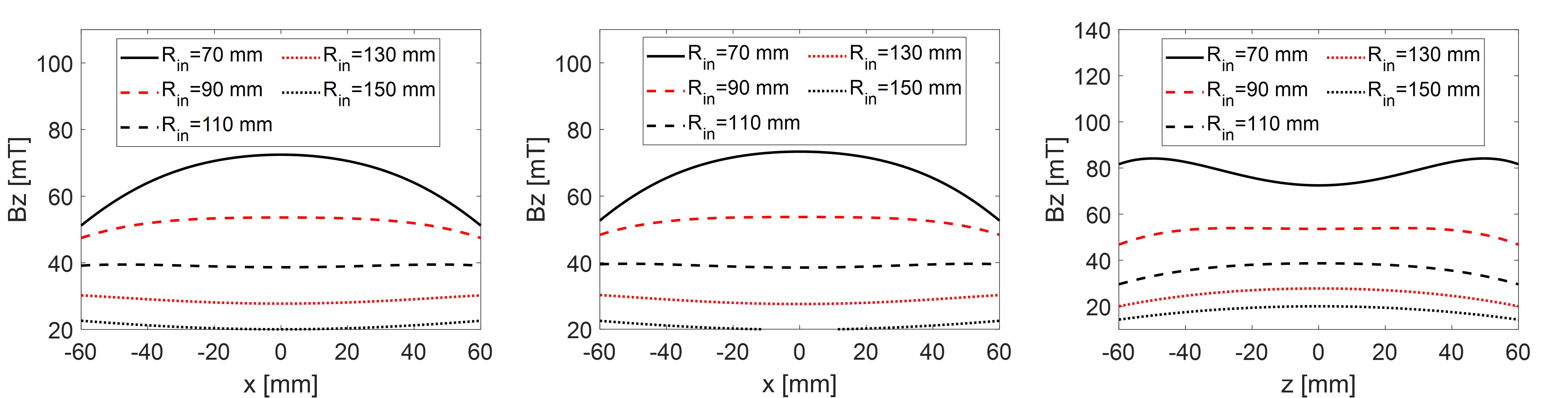}\\
(a)\hspace{0.3\linewidth}(b)\hspace{0.3\linewidth}(c)\\    \includegraphics[width=\linewidth]{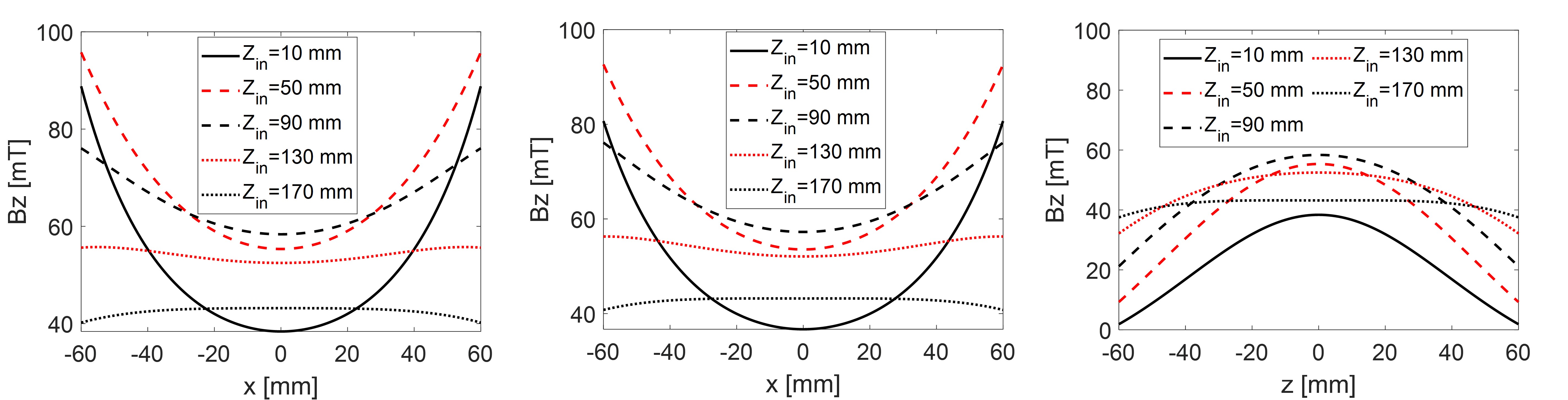}\\
(d)\hspace{0.3\linewidth}(e)\hspace{0.3\linewidth}(f)\\ 	
\includegraphics[width=\linewidth]{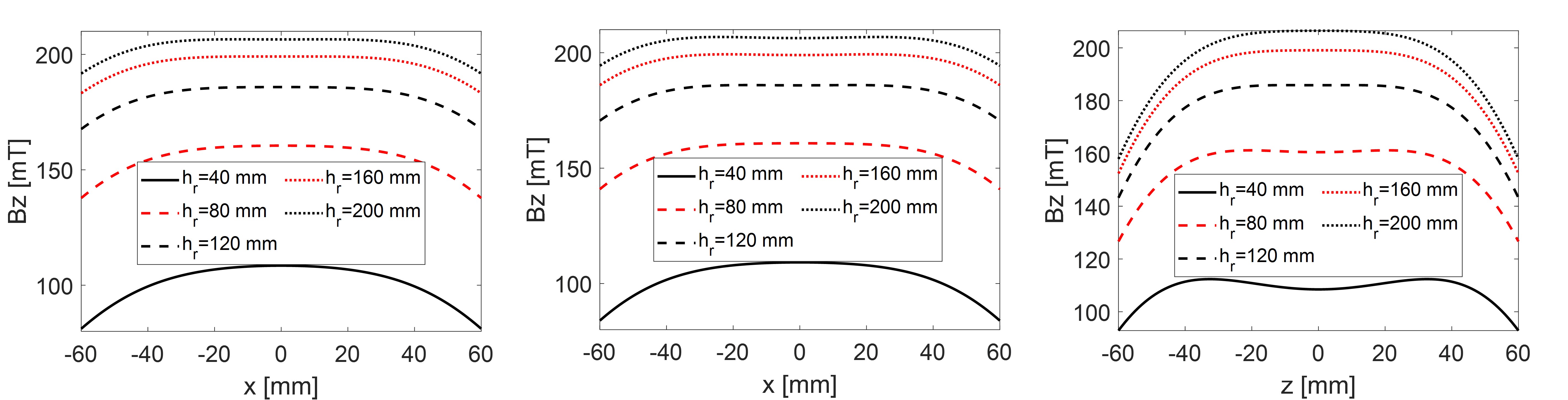}\\
(g)\hspace{0.3\linewidth}(h)\hspace{0.3\linewidth}(i)\\    \includegraphics[width=\linewidth]{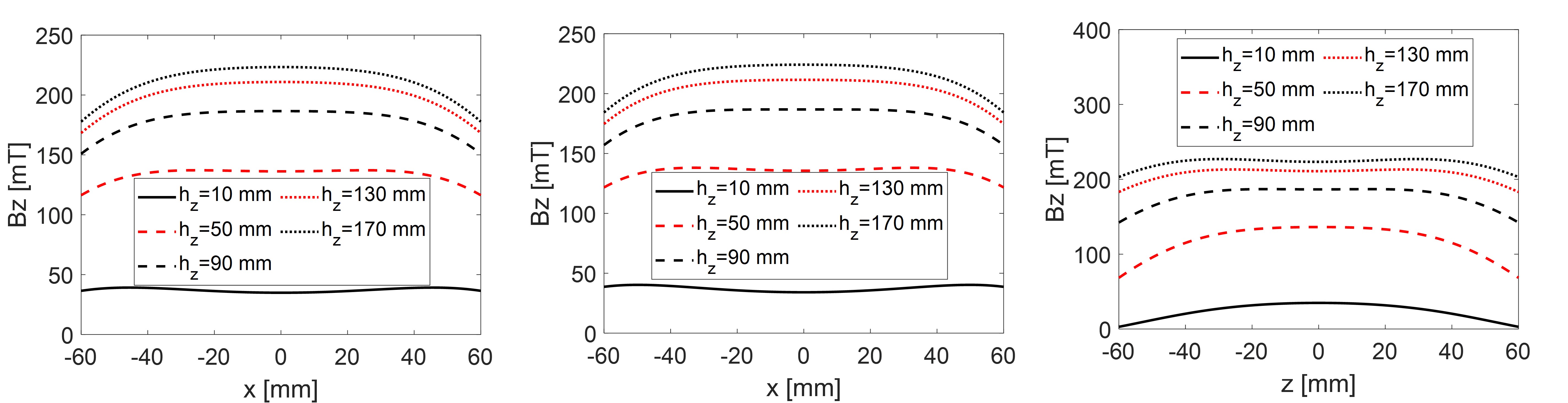}\\
(j)\hspace{0.3\linewidth}(k)\hspace{0.3\linewidth}(l)\\ 	
\end{center}
\caption{Row 1: The $B_\text{z}$ magnetic field component along (a) $x\in[-60,60]$ mm, $y=z=0$, (b) $x\in[-60,60]$ mm, $y=0,z=10$ mm, and (c) $z\in[-60,60]$ mm, $x=y=0$ when sweeping the inner radius $R_\text{in}$ ($Z_\text{in}$=160mm, $h_\text{r}$=40mm, $h_\text{z}$=40mm). Row 2: The $B_\text{z}$ magnetic field component along (d) $x\in[-60,60]$ mm, $y=z=0$, (e) $x\in[-60,60]$ mm, $y=0,z=10$ mm, and (f) $z\in[-60,60]$ mm, $x=y=0$ when sweeping the inner distance $Z_\text{in}$ between the ring pair ($R_\text{in}$=100mm, $h_\text{r}$=40mm, $h_\text{z}$=40mm). Row 3: The $B_\text{z}$ magnetic field component along (g) $x\in[-60,60]$ mm, $y=z=0$, (h) $x\in[-60,60]$ mm, $y=0,z=10$ mm, and (i) $z\in[-60,60]$ mm, $x=y=0$ when sweeping the radial thickness $h_\text{r}$ ($R_\text{in}$=70mm, $Z_\text{in}$=130mm, $h_\text{z}$=50mm). Row 4: The $B_\text{z}$ magnetic field component along (j) $x\in[-60,60]$ mm, $y=z=0$, (k) $x\in[-60,60]$ mm, $y=0,z=10$ mm, and (l) $z\in[-60,60]$ mm, $x=y=0$ when sweeping the axial thickness $h_\text{z}$ ($R_\text{in}$=70mm, $Z_\text{in}$=100mm, $h_\text{r}$=40mm).}
\label{fig:result_four}
\end{figure}

Besides the effects of discretization (i.e., the number of magnet blocks around a ring, $n_\text{bar}$) on the magnetic field patterns, the effects of the other design parameters of an IO ring array, $R_\text{in}$, $Z_\text{in}$, $h_\text{z}$, and $h_\text{r}$ on the field pattern of the array are further investigated. For this part of the study, $n_\text{bar}=16$ and $a=20$\,mm. ``MagTetris'' was used and the FoV stays the same. 

\subsubsection{Effects of $R_\text{in}$}
\label{subsub_Rin}
To examine the effect of $R_\text{in}$, IO ring arrays with $R_\text{in}$ from 70\,mm, 130\,mm at a step of 20\,mm were simulated. The other parameters were set to $Z_\text{in}$\,=\,160\,mm, $h_\text{r}$\,=\,40\,mm, $h_\text{z}$\,=\,40\,mm. The bigger $R_\text{in}$ is, bigger is the bore size which decides which body part or whether a whole body the corresponding system can scan, and the cross-sectional dimensions of the magnet array is bigger. 

The calculated 1D $B_z$s are shown in the first row in Fig.\,\ref{fig:result_four}. As shown in Fig.\,\ref{fig:result_four}\,(a) and (b), when $R_\text{in}$ increases, the field distributions along the $x$-direction (either z=0\,mm or 10\,mm) change from convex downward to flat, then slightly concave upward. This implies homogeneous field distributions can be obtained by varying $R_\text{in}$ for the given $Z_\text{in}$, $h_\text{r}$, $h_\text{z}$, and $n_\text{bar}$. Meanwhile, the homogeneity is obtained at a price of a reduction in field strength. 
Furthermore, the field distribution along the $z$-direction is shown in Fig.\,\ref{fig:result_four}\,(c). As shown, 
at $R_\text{in}$ = 70\,mm, it shows a gradient of 275 mT/m between z\,=\,0\,mm and 40\,mm, whereas at $R_\text{in}$ = 130\,mm, it shows a gradient of 78 mT/m between z\,=\,0\,mm and 40\,mm. When $R_\text{in}$ increases, the change in the field strength becomes less by moving the FoV away from the center along the $z$-direction. 
Combining both observations, when $R_\text{in}$ is set to be right, both field homogeneity on the $xy$-place and a linear gradient along the $z$-direction can be obtained. 

\subsubsection{Effects of $Z_\text{in}$}
\label{subsub_Zin}
To examine how $Z_\text{in}$ effect the field patterns, the IO ring arrays with $Z_\text{in}$ set in the range of 10\,mm to 170\,mm at a step size of 40\,mm were simulated. The rest of the parameters were set to $R_\text{in}$\,=\,100\,mm, $h_\text{r}$\,=\,40\,mm, $h_\text{z}$\,=\,40\,mm. The bigger $Z_\text{in}$ is, bigger is the region to access either for scan or for intervention, and longer is the magnet array.

The calculated 1D $B_z$s are shown in the second row in Fig.\,\ref{fig:result_four}. Fig\,\ref{fig:result_four}\,(d) and (e) shows the field distributions along the $x$-direction at z\,=\,0\,mm and 10\,mm, respectively. As shown, when $Z_\text{in}$ increases from 10\,mm (when two rings are very close to each other) to 170\,mm, the field distribution changes from concave upward to flat at $Z_\text{in}$=130\,mm, then slightly convex downward.
Two observations here. One is that when the two rings are very close to each other, there is a big fluctuation of the fields from the periphery to the center of the FoV. In this case, the $B_z$ fields at the center of the FoV is highly compromised, which is due to the significantly tilted flux away from the $z$-direction when $Z_\text{in}$ is small.
The other observation is that a homogeneous field is possible when $Z_\text{in}$ is set properly with other given design parameters.
At $Z_\text{in}$ = 130\,mm, the IO ring array has a homogeneity of 5.96\% and 7.69\% at z\,=\,0\,mm and 10\,mm, respectively.
Fig.\,\ref{fig:result_four}\,(f) shows the field distributions along the $z$-direction at different $Z_\text{in}$s. As shown, from $Z_\text{in}$ = 10\,mm to 130\,mm, all cases shows monotonic gradients from z\,=\,0\,mm to z\,=\,60\,mm. At $Z_\text{in}$ = 130\,mm when the field is homogeneous, it shows a gradient of 191 mT/m between z=0\,mm and 60\,mm. 

\subsubsection{Effects of $h_\text{r}$}
\label{subsub_hr}
The width of the ring, $h_\text{r}$, also effecting the field pattern of an IO ring array. To examine the effects of $h_\text{r}$, IO ring arrays with $h_\text{r}$ set between 40\,mm to 200\,mm at a step size of 40\,mm were simulated. The rest of the parameters were set to $R_\text{in}$\,=\,70\,mm, $Z_\text{in}$\,=\,130\,mm, $h_\text{z}$\,=\,50\,mm. The corresponding filling factors are from 0.30 to 0.57. When $h_\text{r}$ increases, besides an increased filling factor, the IO ring array becomes bigger and heavier. 

The calculated results are shown in the third row of Fig.\,\ref{fig:result_four}. 
Fig\,\ref{fig:result_four}\,(g) and (h) shows the field distributions along the $x$-direction at z\,=\,0\,mm and 10\,mm, respectively. As shown both in Fig\,\ref{fig:result_four}\,(g) and (h), when $h_\text{r}$ increases from 40\,mm to 200\,mm, the field strengths on the $xy$-plane increase (from 97\,mT to 202\,mT) and becoming more homogeneous (from 28.2\% to 7.4\%). 
Fig.\,\ref{fig:result_four}\,(i) shows the field distribution along the $z$-direction at different $h_\text{r}$s. As shown, at $h_\text{r}$ =40\,mm, there is a gradient of 123.8\,mT/m between z\,=\,0\,mm and $\pm$20\,mm and a higher one of 915.8\,mT/m between z\,=\,$\pm$40\,mm and $\pm$60\,mm. When $h_\text{r}$ increases, only in the region between z\,=\,$\pm$40\,mm and $\pm$60\,mm shows gradients. When $h_\text{r}$ \,=\,200\,mm, the array has an extremely high gradient at 1856.6\,mT/m between z\,=\,$\pm$40\,mm and $\pm$60\,mm. 
Based on the observation, a bigger $h_\text{r}$ (a longer magnet block) can lead to the designs of homogeneous field with a high $z$-gradient. It should be noted that $h_\text{r}$ is along the direction of magnet polarization and the ratios between $h_\text{r}$ and the other two dimension of a block are usually limited by the manufacturing techniques of a magnet manufacturer. Meanwhile, when $h_\text{r}$ increase, the size and weight of the resultant IO ring array both increase, which should become a concern if portability is one of the design goals.


\subsubsection{Effects of $h_\text{z}$}
\label{subsub_hz}
The thickness of the ring, $h_\text{z}$, plays a role in varying the field pattern an IO ring array generates. To understand the effects in detail, IO ring arrays with $h_\text{z}$ set in the range of 10\,mm to 90\,mm at a step size of 40\,mm were simulated. The rest of the parameters were set to be $R_\text{in}$\,=\,70\,mm, $Z_\text{in}$\,=\,100\,mm, $h_\text{r}$\,=\,40\,mm. Bigger the $h_\text{z}$, longer and heavier is the magnet array. 

The calculated results are shown in the fourth row in Fig.\,\ref{fig:result_four}.
Fig\,\ref{fig:result_four}\,(j) and (k) shows the field distributions along the $x$-direction at z\,=\,0\,mm and 10\,mm, respectively. As shown, when $h_\text{z}$ increases, the region with homogeneous fields reduces while field strength increases. 
At $h_\text{z}$\,=\,10\,mm, the fields within a radius of 60\,mm are relatively homogeneous (11.3\% at z\,=\,0\,mm and 16.0\% at z\,=\,10\,mm). The region of having homogeneous field is reduced to a radius of 40\,mm and the field lines become concave downward when $h_\text{z}$\,=\,50\,mm and 90\,mm for both the center slide and that at z\,=\,10\,mm. 
It is also noticed that the increase speed of field strength slows down when $h_\text{z}$ increases, which is because when $h_\text{z}$ increases at a fixed $Z_\text{in}$=100mm, the increase amount of magnet is farther away from the FoV at a higher $h_\text{z}$.  
Fig.\,\ref{fig:result_four}\,(l) shows the field distribution along the $z$-direction at different $Z_\text{in}$s. When $h_\text{z}$ increases, the region that shows gradients becomes from z\,=\,$\pm$20\,mm to $\pm$60\,mm at $h_\text{z}$\,=\,10\,mm, to from $\pm$40\,mm to $\pm$60\,mm at $h_\text{z}$\,=\,170\,mm, with an increase in gradient from 739\,mT/m to 1087\,mT/m. Based on the observations, homogeneous fields on the $xy$-plane can be obtained with high $z$-gradients when $h_\text{z}$ is set properly.

\subsubsection{Further analysis}
Based on the observations and analysis on Fig.\,\ref{fig:result_four}, with a fixed $n_\text{bar}$ and one of the dimensions of a magnet block, $a$, by adjusting one of the other design parameters, $R_\text{in}$, $Z_\text{in}$, $h_\text{z}$, and $h_\text{r}$,  homogeneous fields can be obtained on the $xy$-plane with high $z$-gradients. 
As shown in part\,\ref{subsub_Rin} and \ref{subsub_Rin}, without changing the thickness of the rings in either the radial or the $z$-direction, field homogeneity can be obtained on the $xy$-plane by enlarging the hole of the rings or making the two ring further apart at a price of a reduction at the field strength. On the other hand, As shown in part\,\ref{subsub_hr} and \ref{subsub_hz}, field homogeneity and an increase in field strength on the $xy$-plane can simultaneously be obtained by making the the rings thicker in either the radial or the $z$-direction, at a price of a reduced area. Overall, it strongly suggests that When all the design parameters are optimized together, the chance of achieving high magnetic field with high field homogeneity and high $z$-gradient is high.

\section{Variants of IO ring array}
\label{sec:variants}

\subsection{Asymmetric IO ring array}
\label{subsec:asymmetric_ring}

\begin{figure}[t]
\centering
\begin{center}
\newcommand{\patchSize}{2.45cm}
\scriptsize
\setlength\tabcolsep{0.1cm}
\includegraphics[width=\linewidth]{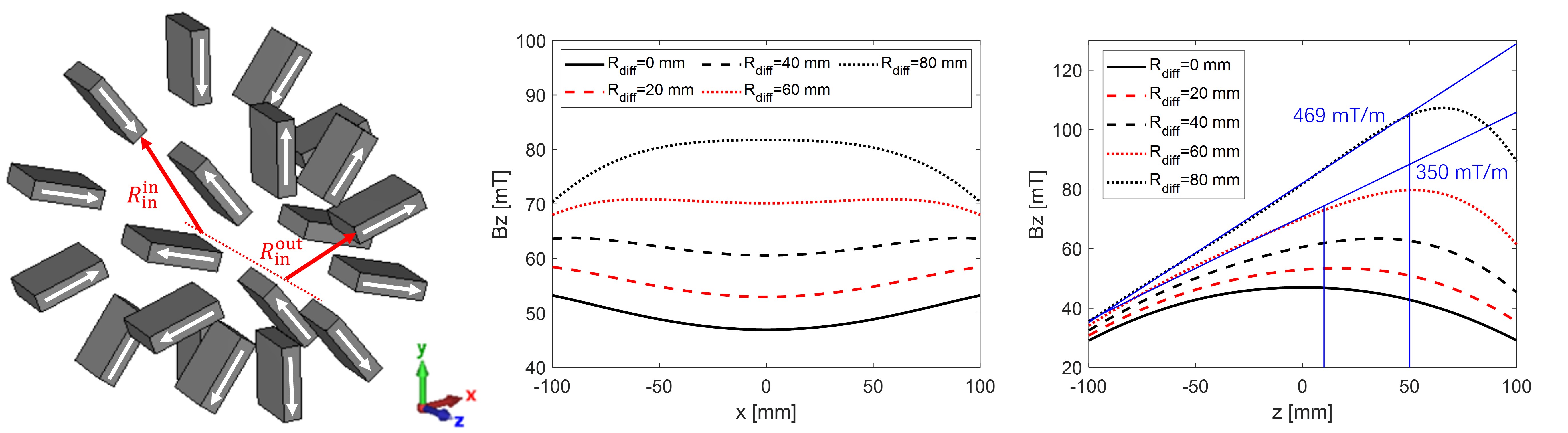}\\
(a)\hspace{0.3\linewidth}(b)\hspace{0.3\linewidth}(c)\\
\end{center}
\caption{(a) The perspective view of an asymmetric IO ring array. The magnetic field along (b) the $x$-axis and (c) the $z$-axis as the difference between the radii of the two rings varies from 0 to 80\,mm. $R_\text{in}=180$ mm for the big ring, and $Z_\text{in}=200$ mm, $h_\text{r}=120$ mm, $h_\text{z}=90$ mm, $a=30$ mm. Linear regression was performed for the cases where $R_\text{diff}=60$ and $80$ mm. The slope value is 350 mT/m for $R_\text{diff}=60$ mm, and 469 mT/m for $R_\text{diff}=80$ mm.}
\label{fig:field_asym_ring}
\end{figure}

One intuitive variant of an IO ring array is the asymmetric IO ring array, where the two rings have individual design parameters. 
Three types of IO ring arrays are studied. 

The first type is when the two rings have different inner radii. Fig.\,\ref{fig:field_asym_ring}\,(a) shows an example. This type of IO ring array was proposed as a helmet-shaped IO ring array in the literature\,\cite{Ren2019_Helmet}. The difference in the radii, $R_\text{diff}$, is defined as $R_\text{diff}=R^\text{in}_\text{in}-R^\text{out}_\text{in}$ where the superscript in/out indicates the inward/outward ring.
At different $R_\text{diff}$s, the field patterns are still concentric. The difference in $R_\text{in}$ affects the field homogeneity on the transversal planes and introduces linear gradients in the longitudinal direction. 

Fig.\ref{fig:field_asym_ring}\,(b) shows the field plots along the $x$-direction at z = 0\,mm when $R^\text{in}_\text{in}=$180\,mm and $R_\text{diff}$ is varies from 0 to 80\,cm. The rest of the parameters were set to $n_\text{bar}=10$, $Z_\text{in}=200$\,mm, $h_\text{r}=120$ mm, $h_\text{z}=90$ mm, and $a=30$. As can be seen, when $R_\text{diff}$ increases and the outward ring becomes smaller, the field along the $x$-axis has the change from convex downward to flat, then concave upward. It suggests that tuning $R_\text{diff}$ has equivalent effect as tuning $R_\text{in}$, which brings more possibilities for the IO ring array geometry under different applications.
Fig.\ref{fig:field_asym_ring}\,(c) shows the field distribution in the $z$-direction within the gap between the two rings at different $R_\text{diff}$s. As shown, the magnetic field along the $z$-axis becomes lopsided with increasing $R_\text{diff}$. It should be noted that when the difference is big enough, for example in this study when $R_\text{diff}\geq 60$ mm, the magnetic field exhibits linear gradient across a considerably long distance along the $z$-direction. As shown in Fig.\ref{fig:field_asym_ring}\,(c), at $R_\text{diff}=60$\,mm, it shows a gradient of 350\,mT/m from -100\,mm to 10\,mm whereas at $R_\text{diff}=80$\,mm, it shows a gradient of 469\,mT/m from -100\,mm to 50\,mm.
It is noted that in the case when $R^\text{out}_\text{in}=$180\,mm and $R_\text{diff}$ is varies from -80 to 0\,cm, the changes in field patterns along the $x$-axis when $R_\text{diff}$ changes will be identical, and those along the $z$-axis will be mirror symmetric with respect to the vertical line at z=0\,mm.

Combining the observations in Fig.\ref{fig:field_asym_ring}\,(b) and (c), an asymmetric IO ring array can be designed with high field homogeneity and strength, and a linear gradient along the $z$-direction. 
Comparing to the effect by tuning $R_\text{in}$ to obtain a field homogeneity, with comparable values of other design parameters, a higher field strength is possible by designing this type of asymmetric IO ring array.


\begin{figure}[t]
\centering
\begin{center}
\newcommand{\patchSize}{2.45cm}
\scriptsize
\setlength\tabcolsep{0.1cm}
\includegraphics[width=\linewidth]{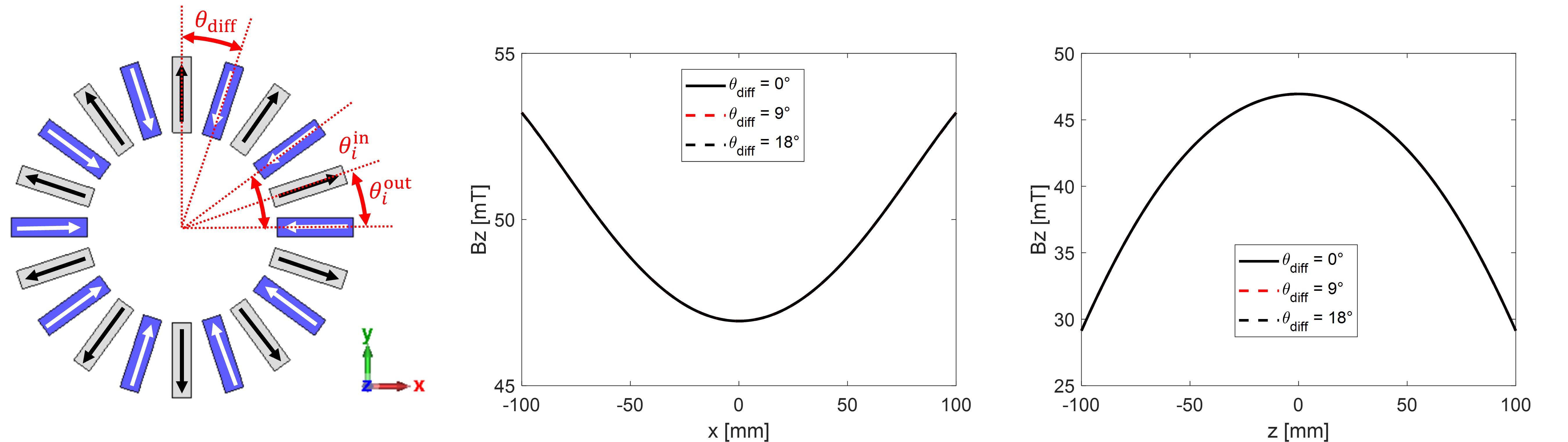}\\
(a)\hspace{0.3\linewidth}(b)\hspace{0.3\linewidth}(c)\\
\end{center}
\caption{(a) An IO ring array with one ring rotated by an angle $\theta_\text{diff}=18\degree$. The $B_\text{z}$ field when sweeping $\theta_\text{diff}$ along (b) the $x$-axis with $x\in[-100,100]$ mm and (c) the $z$-axis with $z\in[-100,100]$ mm. The other design parameters are $n_\text{bar}=10$, $Z_\text{in}=200$\,mm, $h_\text{r}=120$ mm, $h_\text{z}=90$ mm, and $a=30$.}
\label{fig:asym_angle}
\end{figure}

The second type of asymmetry is rotating one ring about the $z$-axis by an angle $\theta_\text{diff}$ in the azimuth direction to misalign the inward and outward rings. 
Fig.\,\ref{fig:asym_angle}\,(a) shows the side view of an example when $n_\text{bar}=10$. As shown, the misalign angle, $\theta_\text{diff} = |\theta^\text{out}_\text{i}-\theta^\text{in}_\text{i}|$, where the subscript $i$ indicates the i$^\text{th}$ magnet in an inward/outward ring. 
The cases with different $\theta_\text{diff}$s are examined. The other design parameters were set as follows, $n_\text{bar}=10$, $Z_\text{in}=200$\,mm, $h_\text{r}=120$ mm, $h_\text{z}=90$ mm, and $a=30$. The angle between two successive magnet blocks in a ring is 36$\degree$. 
Fig.\,\ref{fig:asym_angle}\,(b) and (c) shows the calculated $B_\text{z}$ along the $x$- and $z$-axis, respectively, when $\theta_\text{diff}$ is changed from 0$\degree$ to 18$\degree$ at a step of 9$\degree$. As shown, the misalignment does not change the field distribution in either the $x$- or the $y$-direction much. The different can be more obvious when $n_\text{bar}$ is small. 


\begin{figure}[t]
\centering
\begin{center}
\newcommand{\patchSize}{2.45cm}
\scriptsize
\setlength\tabcolsep{0.1cm}
\includegraphics[width=\linewidth]{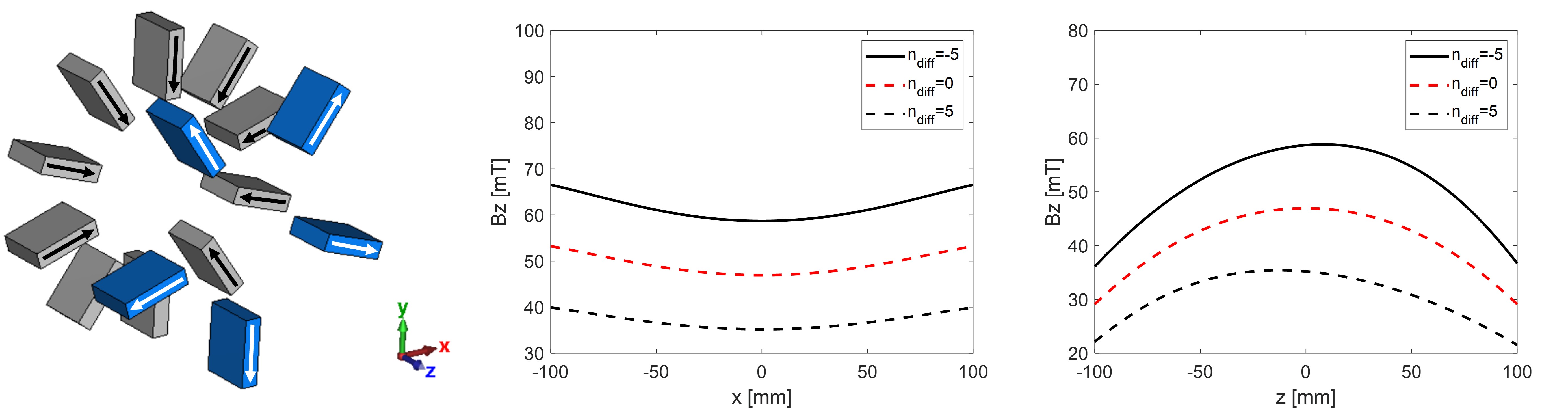}\\
(a)\hspace{0.3\linewidth}(b)\hspace{0.3\linewidth}(c)\\
\end{center}
\caption{(a) An IO ring array with one ring has the number of block as $n_\text{bar}'=n_\text{bar}-n_\text{diff}=5$. The $B_\text{z}$ field when sweeping $n\text{diff}$ along (b) the $x$-axis with $x\in[-100,100]$ mm and (c) the $z$-axis with $z\in[-100,100]$ mm. The other design parameters are $n_\text{bar}=10$, $Z_\text{in}=200$\,mm, $h_\text{r}=120$ mm, $h_\text{z}=90$ mm, and $a=30$.}
\label{fig:asym_n_bar}
\end{figure}

The third type of asymmetric IO ring array to be discussed here is one where the inward and outward ring have different $n_\text{bar}$. Fig.\,\ref{fig:asym_n_bar}\,(a) shows an example when the inward ring has 10 magnet bars, $n^\text{in}_\text{bar}$\,=\,10, and the outward ring has 5 magnet bars, $n^\text{out}_\text{bar}$\,=\,5. The difference in the numbers of inward and outward magnets, $n_\text{diff}$, is defined as $n_\text{diff}=n^\text{in}_\text{bar}-n^\text{out}_\text{bar}$. 
The cases with different $n_\text{diff}$s were studied. The other design parameters were set as follows, $Z_\text{in}=200$\,mm, $h_\text{r}=120$ mm, $h_\text{z}=90$ mm, and $a=30$.
Fig.\,\ref{fig:asym_n_bar}\,(b) and (c) shows 
the corresponding $B_\text{z}$ along the $x$- and $z$-axis when $n_\text{diff}=-5,0,5$ (i.e., $n^\text{out}$=15, 10, 5), respectively. 
It is observed that the difference in $n_\text{bar}$ does not change the shape of the distribution curve significantly, i.e., $n_\text{diff}$ has negligible effects on the field homogeneity. For field strengths, the case with the more magnet blocks has a higher field strength.  


\subsection{Partial IO ring array}
\label{subsec:partial_ring}

\begin{figure}[t]
\centering
\begin{center}
\newcommand{\patchSize}{2.45cm}
\scriptsize
\setlength\tabcolsep{0.1cm}
\includegraphics[width=\linewidth]{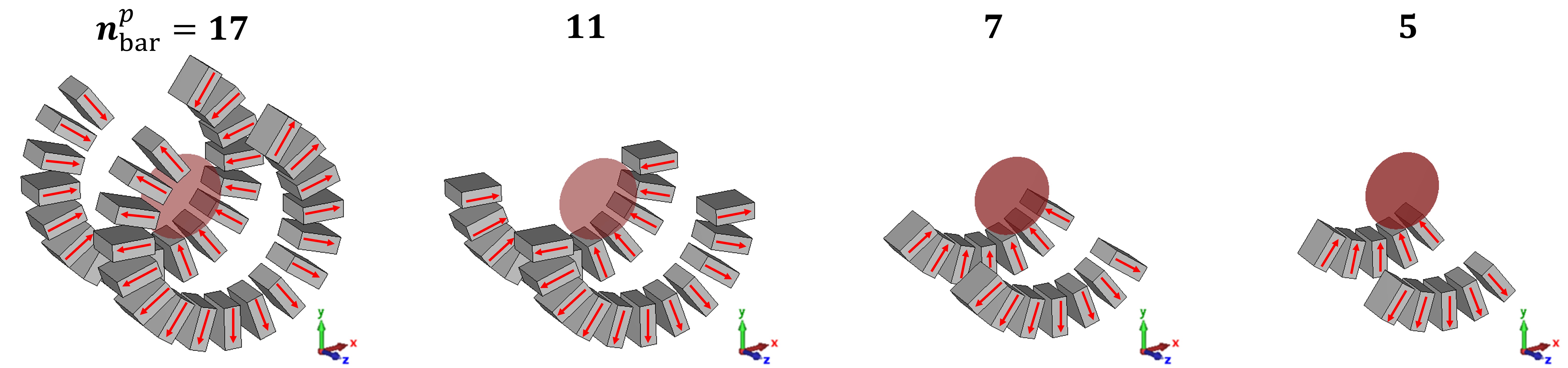}\\
(a)\hspace{0.25\linewidth}(b)\hspace{0.25\linewidth}(c)\hspace{0.25\linewidth}(d)\\
\includegraphics[width=\linewidth]{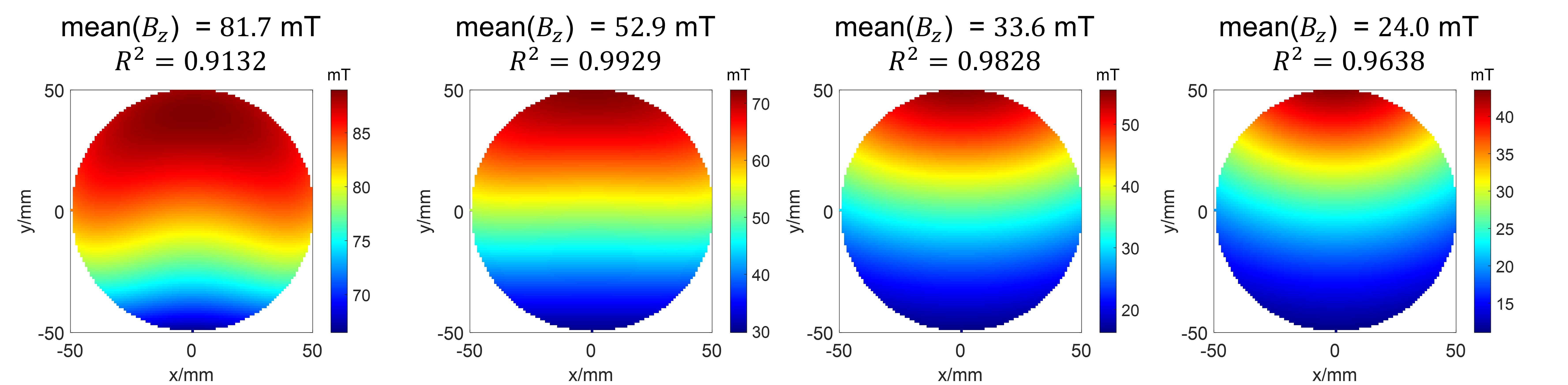}\\
(e)\hspace{0.25\linewidth}(f)\hspace{0.25\linewidth}(g)\hspace{0.25\linewidth}(h)\\
\end{center}
\caption{The magnetic field by part of the IO ring array as shown in lower-right sub-figure. Row 1: The side view of a partial IO ring with $n_\text{bar}=17,11,7,5$. Row 2: magnetic field pattern on the $z=0$ with an FoV radius of 50\,mm in the corresponding case. The design parameters are $R_\text{in}=90$ mm, $Z_\text{in}=150$ mm, $h_\text{r}=h_\text{z}=50$ mm, and $a=20$ mm. The angle between neighboring magnets is $18\degree$}
\label{fig:field_partial_ring}
\end{figure}
\begin{figure}[t]
\centering
\begin{center}
\newcommand{\patchSize}{2.45cm}
\scriptsize
\setlength\tabcolsep{0.1cm}
\includegraphics[width=0.5\linewidth]{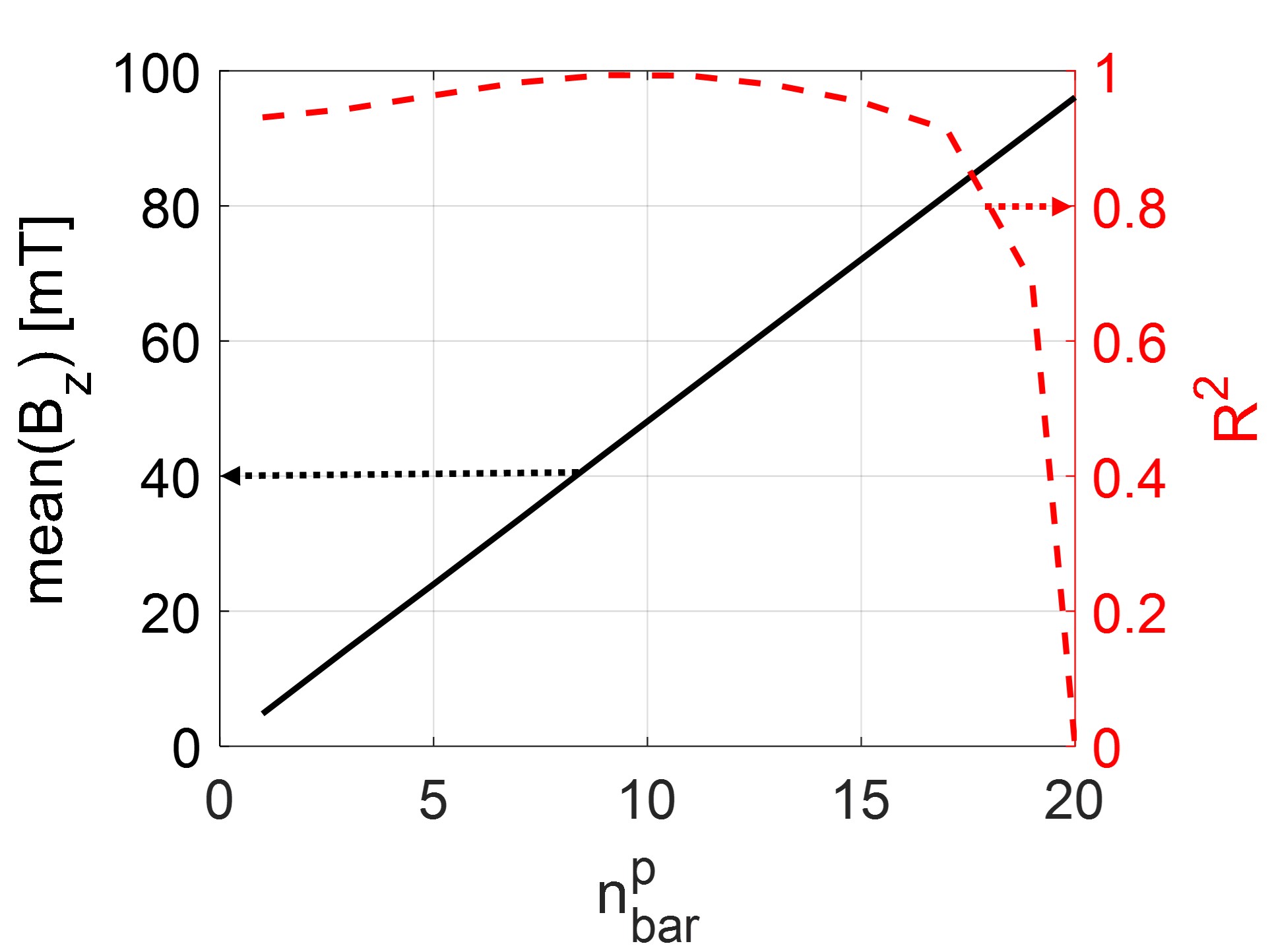}\\
\end{center}
\caption{The average field strength and the linearity of $B_\text{z}$ on the $xy$-plane with a radius of 50\,mm for the partial IO ring array described in Fig.\,\ref{fig:field_partial_ring} with $n_\text{bar}^\text{p}\in[1,20]$.}
\label{fig:partial_ring_n_bar}
\end{figure}

Another variant of an IO ring array is a partial IO ring array when a number of successive magnet blocks are removed, allowing an opening in the cylindrical magnet array. The first row in Fig.\,\ref{fig:field_partial_ring} shows some examples when an IO ring array ($R_\text{in}=90$ mm, $Z_\text{in}=150$ mm, $h_\text{r}=h_\text{z}=50$ mm, $n_\text{bar}=20$) is transformed to partial IO ring arrays with $n^\text{p}_\text{bar}=17,11,7,5$ where the superscript p here stands for partial. When $n^\text{p}_\text{bar}$ decreases, the size of opening increases. 
The opening allows the access to the scanner in the $y$-direction besides that in the $z$-direction. 
When $n^\text{p}_\text{bar}=11$, the PMA becomes an half IO ring array. When $n^\text{p}_\text{bar}<11$, the array becomes single-sided and \textit{ex-situ}, i.e., the FoV is at one side of the magnet array.


The size of the opening affect the field pattern of a partial IO ring array. In this study, the magnetic fields of the structures in Fig.\,\ref{fig:field_partial_ring} (a) - (d) were calculated. Fig.\,\ref{fig:field_partial_ring} (e) - (h) show the calculated fields on the $xy$-plane at z = 0\,mm. 
As shown, with $n^\text{p}_\text{bar}$ decreases, the resulting magnetic fields show monotonic gradients along the $y$-direction and a decrease in the average field strength. When it becomes a semi-circle at $n^\text{p}_\text{bar}=11$, the gradient shows the highest linearity with a linear regression coefficient, $R^2=0.993$. 
The cases when $n^\text{p}_\text{bar}=1$ to a full IO ring array were simulated for further investigations. Fig.\,\ref{fig:partial_ring_n_bar} shows the trend of the average field strength (black solid curve and the vertical axis on the left) and the linearity regression coefficient $R^2$ (dashed red curve and the vertical axis on the right) of the field distribution on the $xy$-plane at z = 0\,mm when $n^\text{p}_\text{bar}$ increases. As shown by the black solid curve, the average field strength grows linearly with an increased $n^\text{p}_\text{bar}$. 
For the linearity of the field pattern, as shown by the red dashed curve, the linearity hit a peak at $n^\text{p}_\text{bar}=11$ when the array is an half IO ring. When $n^\text{p}_\text{bar}\geq17$, $R^2$ starts to drop rapidly. At $n^\text{p}_\text{bar}=20$ when the array is a full IO ring array, the pattern becomes concentric which does not have linearity along the $y$-direction.

A partial IO ring array offers flexible access of the object under scan to the scanner while the field strength is compromised because of a reduced amount of magnet and the field pattern is not homogeneous. Good linearity can be obtained when the opening is around 180\degree. Such a partial IO ring array was proposed for spine imaging\,\cite{liang2023ismrm} in the literature.

\subsection{Convertible IO ring}
\label{subsec:openable_ring}

\begin{figure}[t]
\centering
\begin{center}
\newcommand{\patchSize}{2.45cm}
\scriptsize
\setlength\tabcolsep{0.1cm}
\includegraphics[width=0.8\linewidth]{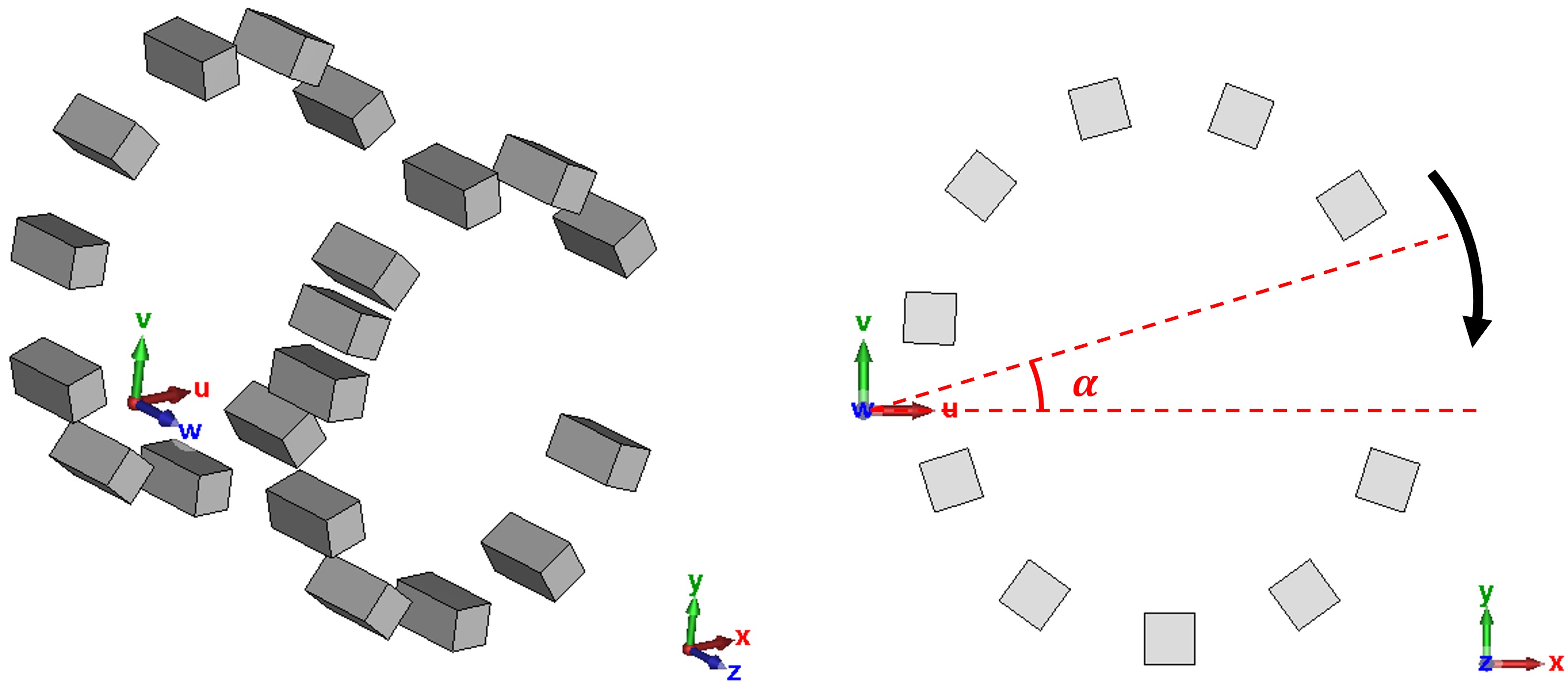}\\
(a)\hspace{0.45\linewidth}(b)\\
\end{center}
\caption{The 3D view and (b) The front view of the convertible IO ring opened for 15\degree about the origin of the local $uvw$-coordinates. The design parameters are $R_\text{in}=80$ mm, $Z_\text{in}=135$ mm, $h_\text{r}=20$ mm, $h_\text{z}=40$ mm, $a=20$ mm, and $n_\text{bar}=10$. The magnetic remanence is assumed to be $B_\text{r}=1.43$ T for the torque calculation.}
\label{fig:openable_ring}
\end{figure}

An IO ring array can be convertible as shown in Fig.\,\ref{fig:openable_ring}. It can offer strong and homogeneous field and allows a flexible access of an object under scan when it is open. This can be important to the scan of the neck\,\cite{ren2017convertible} or a tree trunk of a live tree\,\cite{ren2017convertible,WINDT201127}. 


As shown in Fig.\,\ref{fig:openable_ring}, a pivot is picked between two neighboring magnets (indicated by the origin of a local coordinate, the uvw-coordinates), and the IO ring can be opened around this pivot with two halves separated. Since the IO ring is axially symmetric, the choice of pivot does not affect the force to overcome during the opening and the closure of the convertible PMA.

For the IO ring array described above, the torque during the opening process was calculated. The design parameters are $R_\text{in}=80$ mm, $Z_\text{in}=135$ mm, $h_\text{r}=20$ mm, $h_\text{z}=40$ mm, $a=20$ mm, $n_\text{bar}=10$, and $B_\text{r}=1.43$ T. 
The pivot was set as $(x,y,z)=(-120,0,0)$. When the angle of opening is about 50\degree, the torque to overcome reaches the maximum, which is 61.1 N/m. As a comparison, a Halbach ring array with the same design parameters were calculated. The Halbach array was placed so that the torque during opening is minimized with the same pivot location. When the angle of opening is about 40\degree, the torque to overcome reaches the maximum, which is 157.6 N/m. Therefore, for the same geometry, the IO ring array requires less torque to open the array, which makes the convertible design more feasible and safer. It is noted that an actual Halbach array has multiple rings forming a cylinder rather than two rings that are far apart, and thus the torque for the same angle of opening is expected to be much higher than the current setup under comparison. 

\subsection{IO ring with ferromagnetic yoke}
\label{subsec:with_yoke}

\begin{figure}[t]
\centering
\begin{center}
\newcommand{\patchSize}{2.45cm}
\scriptsize
\setlength\tabcolsep{0.1cm}
\includegraphics[width=0.7\linewidth]{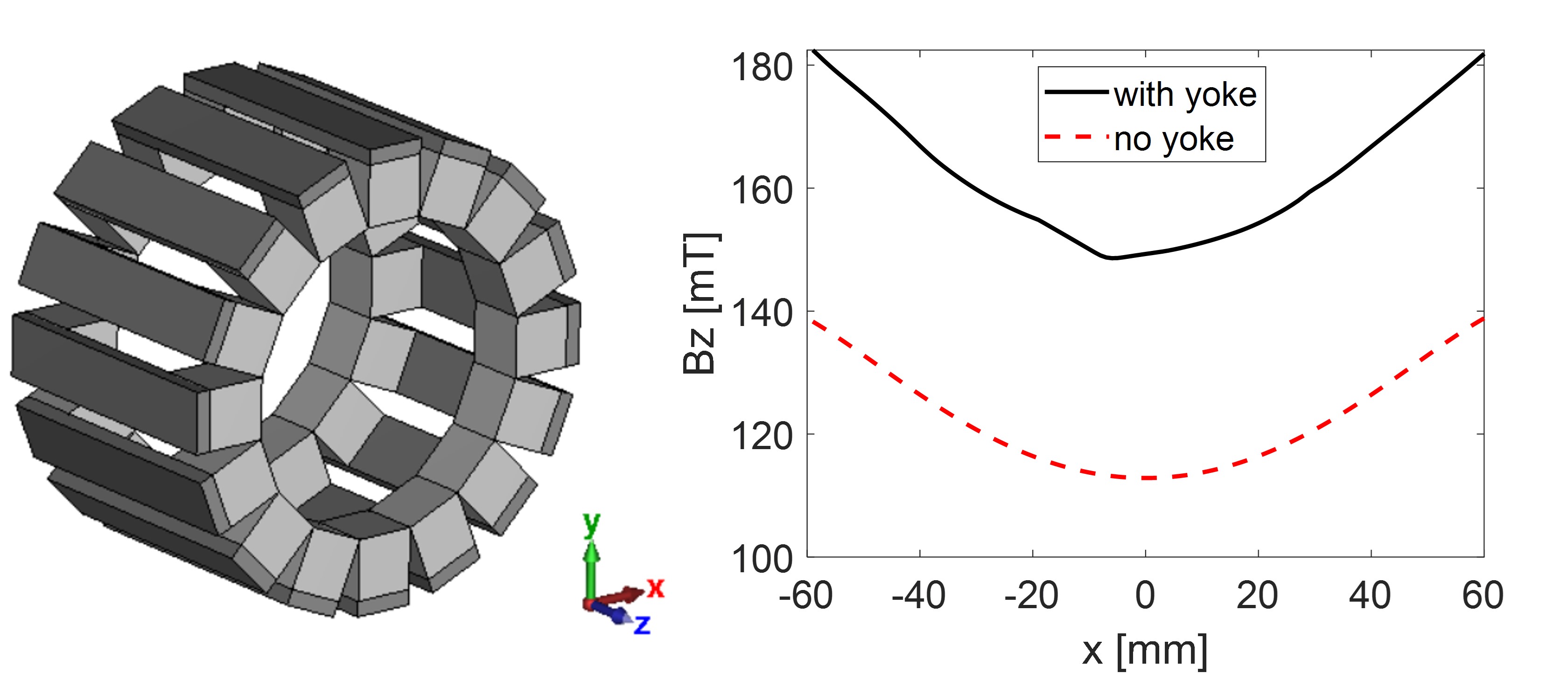}\\
(a)\hspace{0.3\linewidth}(b)\\
\includegraphics[width=0.7\linewidth]{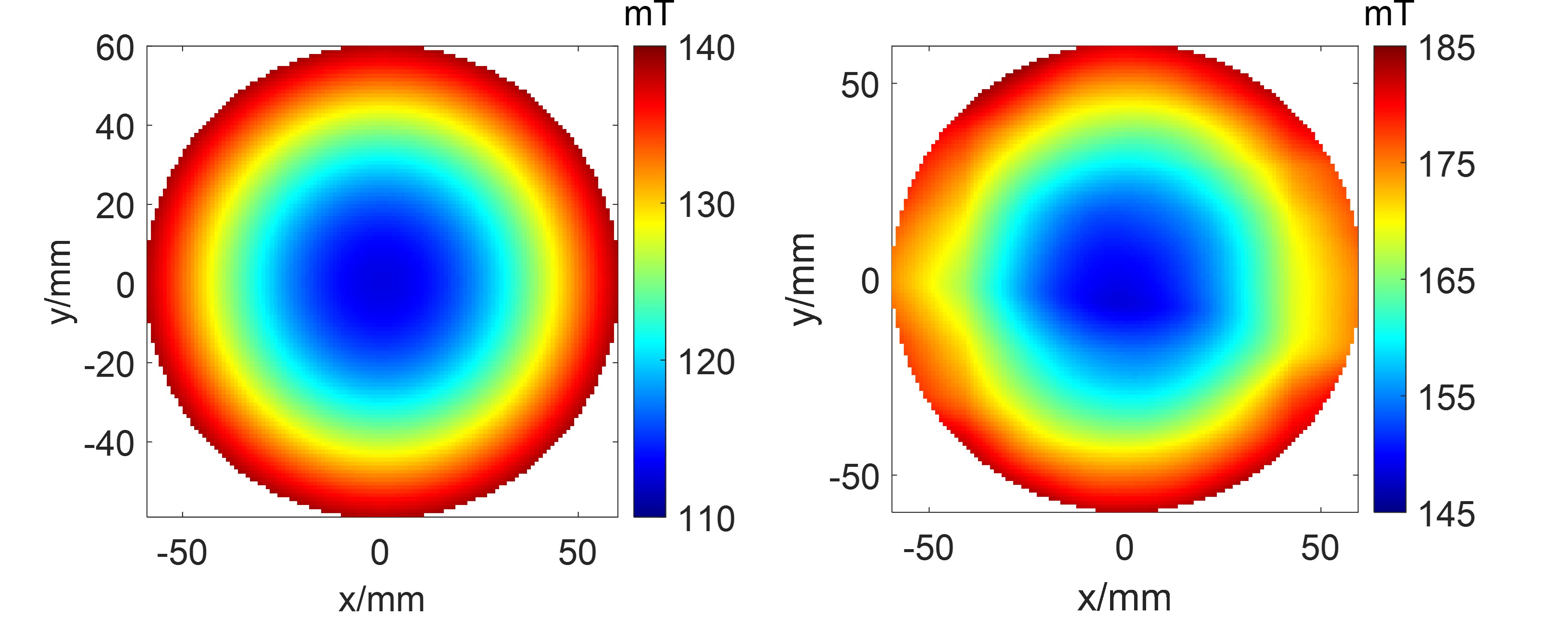}\\
(c)\hspace{0.3\linewidth}(d)\\
\end{center}
\caption{(a) An IO ring array with ferromagnetic yokes connecting the magnet pairs. The design parameters are the same as Fig.\,\ref{fig:result_nbar} with $n_\text{bar}=16$. (b) The magnetic field along the $x$-axis for the cases with/without the iron yokes. (c) The magnetic field on the $xy$-plane calculated by ``MagTetris'' without the iron yokes. (d) The magnetic field on the $xy$-plane simulated by CST with the iron yokes.}
\label{fig:yoke_all}
\end{figure}

Ferromagnetic yokes can be added between a magnet pair to guide the magnetic flux. They can be added to all or some of the magnet pairs. Fig.\,\ref{fig:yoke_all}\,(a) shows an example when all the pairs are connected using ferromagnetic yokes. With the ferromagnetic yoke, the magnet fluxes are guided from the outward magnet through the yoke to the inward one. This leads to a strengthened field inside the IO ring array, a distortion of the field pattern, and a significant reduction in the fringing field. Fig.\,\ref{fig:yoke_all}\,(b) shows the magnetic field along the $x$-axis with/without the iron yokes. It can be seen that adding the iron yokes will improve the field strength within the bore without changing the field pattern. Fig.\,\ref{fig:yoke_all}\,(c) is the same figure as Fig.\,\ref{fig:result_nbar}\,(a), which is the magnetic field without the iron yokes for comparison. Fig.\,\ref{fig:yoke_all}\,(d) shows the field distribution on the $xy$-plane at z=0\,mm of the IO ring array in Section\,\ref{subsec:discretize} at $n_\text{bar}=16$ when iron yokes with a width of $a$, a length of $z_\text{in}+2h_\text{z}$, and a thickness of 10\,mm. The iron yokes are taped on the magnets. Comparing the field distribution in Fig.\,\ref{fig:yoke_all}\,(c) to that in Fig.\,\ref{fig:yoke_all}\,(d), it can be seen that the iron yokes enhances the magnetic field within the bore for about 40\,mT, and the range of the magnetic field remains, which results in higher homogeneity for the case with iron yokes. It suggests that the ferromagnetic yokes helps to confine the magnetic flux within the bore of the IO ring array.

\section{Construction of an IO ring array in the lab}
\label{sec:lab_IO_construct}

\begin{figure}[thp]
\centering
\begin{center}
\newcommand{\patchSize}{2.45cm}
\scriptsize
\setlength\tabcolsep{0.1cm}
\includegraphics[width=0.6\linewidth]{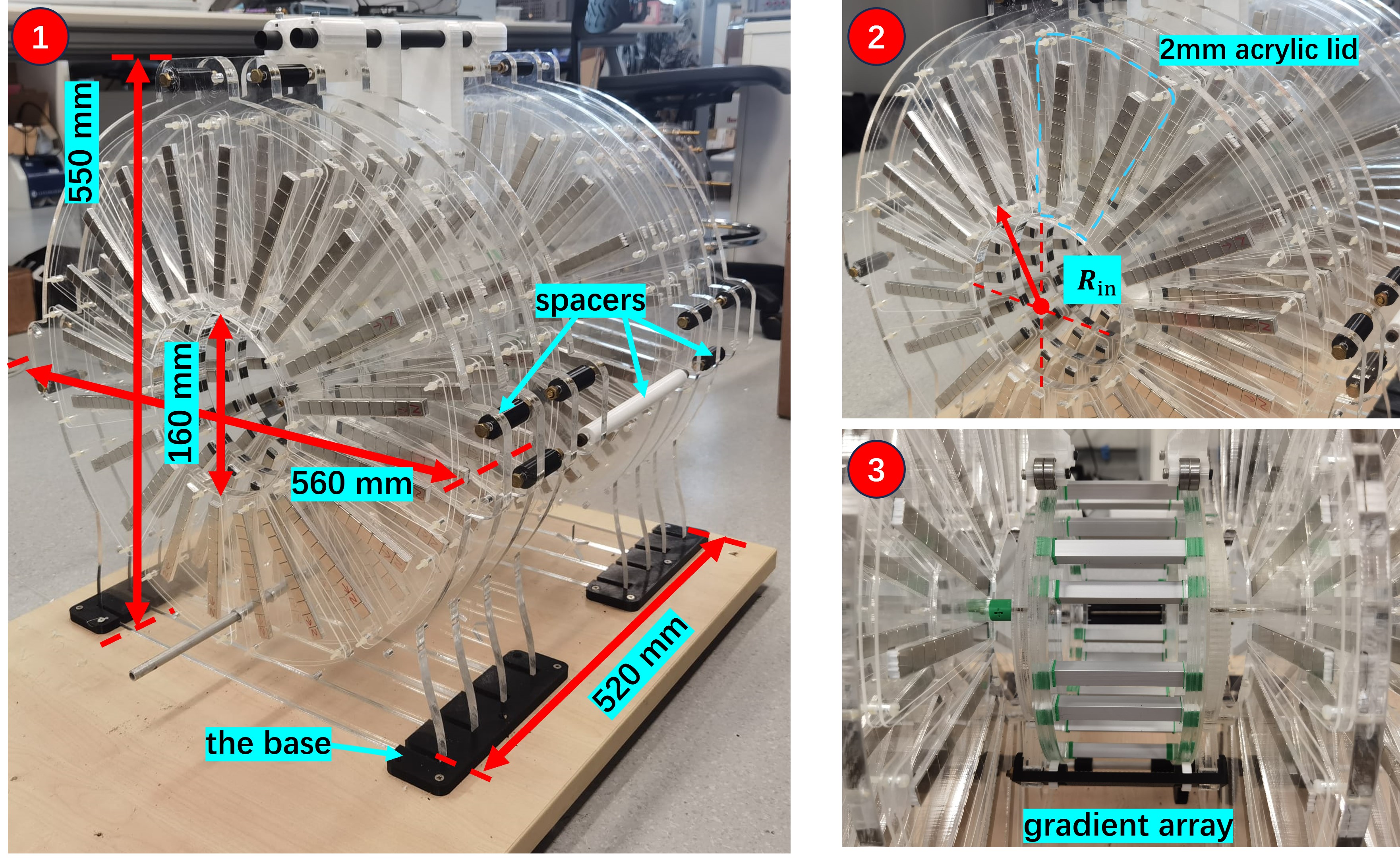}\\
(a)\\
\includegraphics[width=0.6\linewidth]{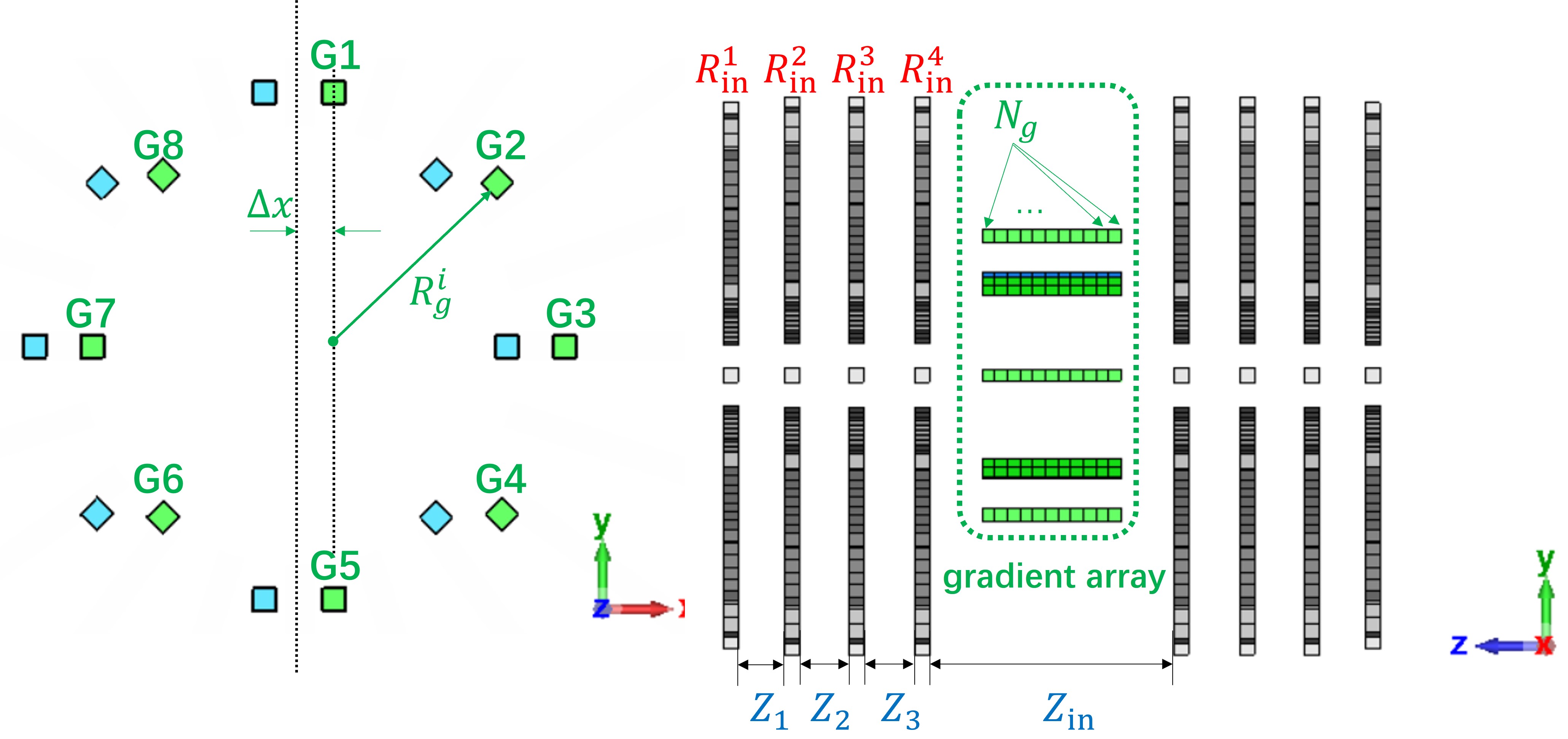}\\
(b)\hspace{0.3\linewidth}(c)\hspace{0.3\linewidth}\\
\includegraphics[width=\linewidth]{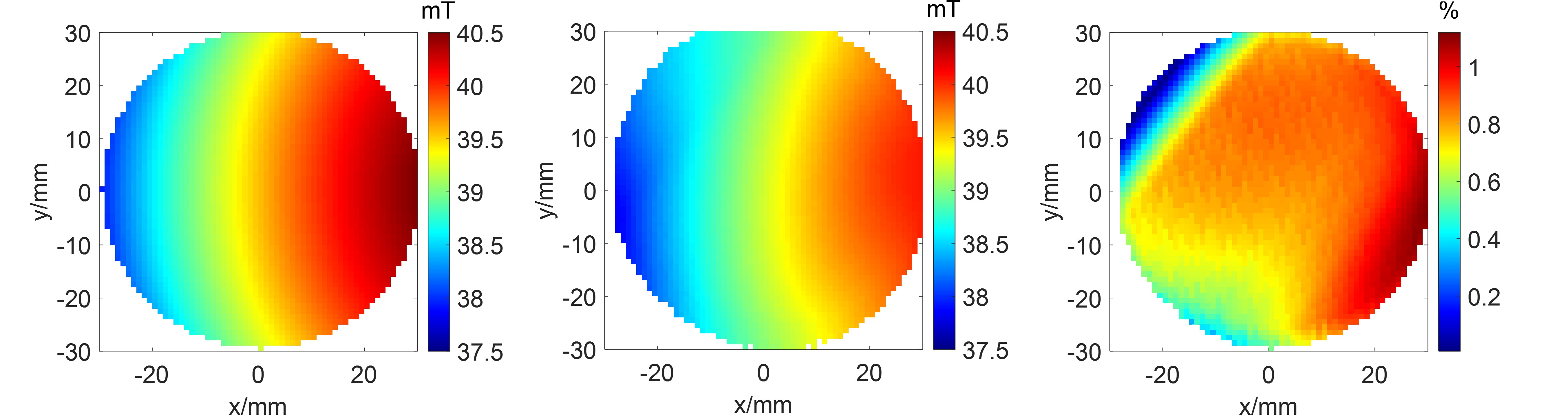}\\
(d)\hspace{0.3\linewidth}(e)\hspace{0.3\linewidth}(f)\\
\end{center}
\caption{\textbf{Row 1}: (a) Sub-figure (1): the overall assembly, (2): the detail for one layer, and (3): the side view showing the gradient array of the IO ring array assembly. \textbf{Row 2}: (b) front view (gradient-array-only) The number of magnets in a column is $N_\text{g}=11$. The offset $\Delta x=15$ mm. The inner radius $R_{g}^i$ for columns $G1-G8$ are $105, 100, 95, 100, 105, 100, 95, 100$\,mm. The blue and green arrays are symmetric about $y$-axis, where the green array has +$z$-polarization, and the blue array has -$z$-polarization. (c) The side view of the IO ring built in the lab. \textbf{Row 3}: The (d) ``MagTetris'' calculation, (e) measurement, and (f) the percentage error plot. The measured plane has a diameter of 60\,mm at z = 0\,mm.}
\label{fig:labIO_all}
\end{figure}

An IO ring array PMA can be built by researchers in a lab when the sizes of the magnets are constrained. The sub-figure\,(1) of Fig.\,\ref{fig:labIO_all}\,(a) shows an in-house built IO ring array PMA designed for a FoV of $\phi 80\times 70$ mm$^3$.  
has a footprint of $560\times550\times520$ mm$^3$, with a 5-Gauss region of $1140\times1140\times1500$ mm$^3$.
It consists of 1600 12$^3$\,mm$^3$ N52 NdFeB magnets ($B_\text{r}=1.43$ T) and 176 10$^3$\,mm$^3$ N45 NdFeB magnets ($B_\text{r}=1.32$ T). 

A magnet block in the IO ring array PMA is implemented by four magnet columns where each column has 10 12\,mm$\times$12\,mm N52 NdFeB magnet cubes and the distances between the columns along the $z$-direction, $Z_1$\,=\,36\,mm, $Z_2$\,=\,39\,mm, and $Z_3$\,=\,40\,mm as shown in Fig.\,\ref{fig:labIO_all}\,(d), while $Z_\text{in}=192$ mm. Therefore, the inner ring or outer ring can be implemented by four layers each of which radially populated with magnet columns. To further optimize the ring disk, the inner radius of the layer ($R^i_\text{in}$, i = 1, 2, 3, and 4) was varied slightly to be 95\,mm for the outermost one and 100\,mm for the rest. 
As shown in the sub-figure\,(3) of Fig.\,\ref{fig:labIO_all}\,(a), the gradient array is implemented using two 12-mm acrylic disks and supporting structures fixed onto the based array. There are 16 aluminum profiles going across the two acrylic disks that contains the magnets. Two 2-mm acrylic boards are fixed at the outer surface of the 12-mm disks to make the aluminum profiles stable.

The magnet columns for each layer are housed by a laser cut acrylic board, forming a ring disk.
The sub-figure\,(2) of Fig.\,\ref{fig:labIO_all}\,(a) shows the detailed structure of a ring disk which consists of a 12-mm thick acrylic board with etched slots where magnet columns are filled and 2-mm arch-shaped acrylic board on both sides of the layer using M3 nylon screws to fix the magnet columns. Each 2-mm arch-shaped acrylic board covers two columns of magnets. The ring disks were built before they are slotted into the base. The slots on the base are formed by 3D printed structures screwed onto a wooden board. 3D printed spacers are added between two successive layers to maintain the distance between layers. With the housing, the magnet has a clear bore size of 160\,mm in diameter. 
Besides the IO ring arrays, it has a gradient array as shown in Fig.\,\ref{fig:labIO_all}\,(b) for encoding\,\cite{LIANG2022IORing}. 
The detailed dimensions of the built IO ring array are included in the caption of Fig.\,\ref{fig:labIO_all}. 

The built IO ring array was both simulated using ``MagTetris'' and characterized experimentally. The measurement was done by a Gaussmeter probe (LakeShore460, resolution = 0.001\,mG) fixed on a xy$z$-moving platform (Physik Instrumente). Fig.\,\ref{fig:labIO_all}\,(d)-(e) show the calculated and the measured magnetic fields in a circular region with a diameter of 60\,mm at z = 0\,mm, respectively, and Fig.\,\ref{fig:labIO_all}\,(f) shows the differences between them. 
As shown, the IO ring array has an average magnetic field strength at 39.7\,mT, a built-in gradient of 4.57\,mT/m (an RF bandwidth of 9.22\,\%), and an average linear regression coefficient of 0.9725. The measured result are close to the simulated one. As shown in Fig.\,\ref{fig:labIO_all}\,(f), the field map of the in-house built IO ring array shows discrepancies within 1\% compared to the simulated one, indicating good precision of magnet assembly. 

\begin{figure}[t]
\centering
\begin{center}
\newcommand{\patchSize}{2.45cm}
\scriptsize
\setlength\tabcolsep{0.1cm}
\includegraphics[width=0.6\linewidth]{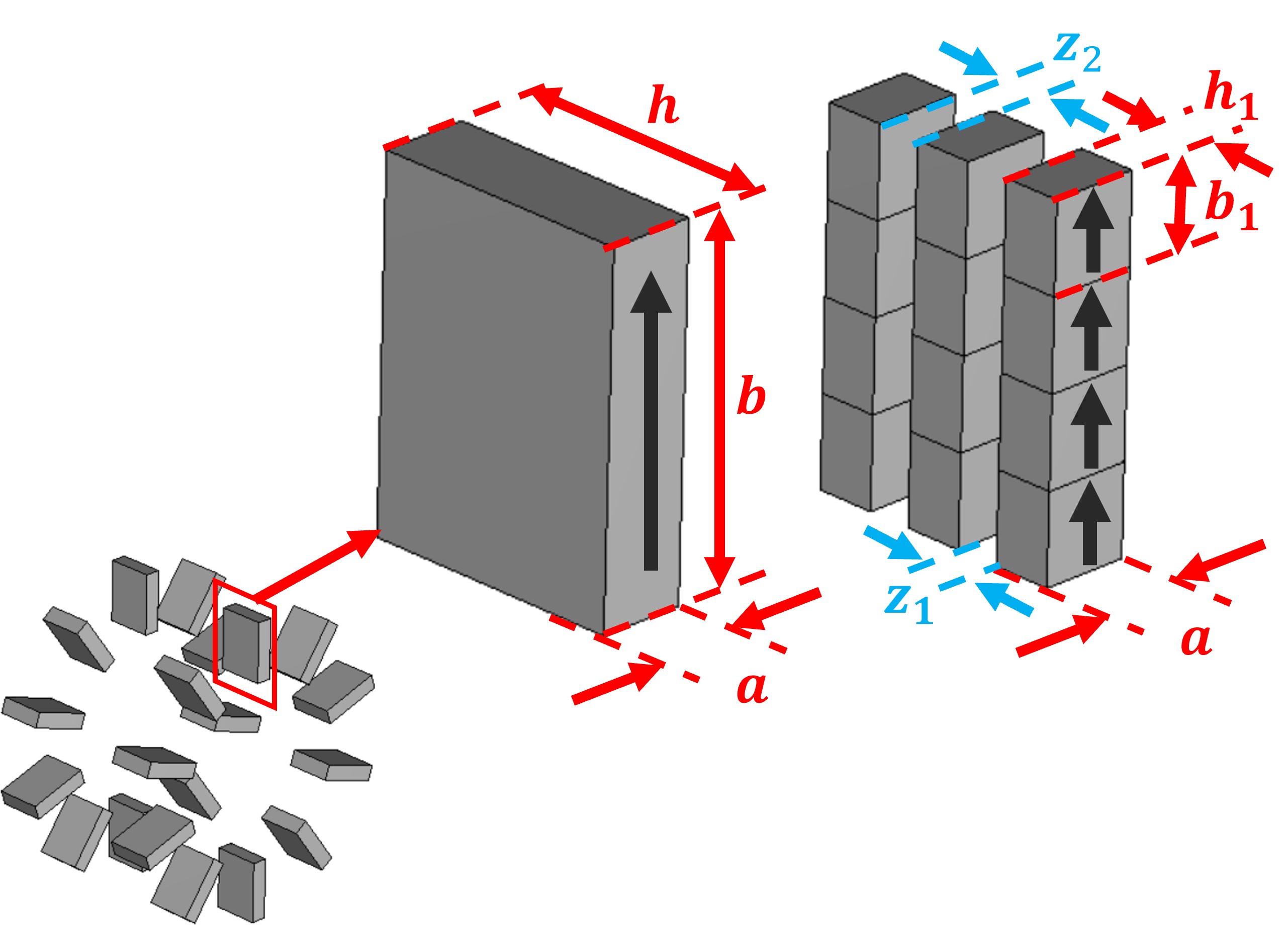}\\
\end{center}
\caption{A demonstration of discretizing the big magnet block for IO ring. For practical assembly, a big magnet block will be replaced by several layers of small magnet blocks, while there is enough spacing between layers to reduce the repulsion.}
\label{fig:labIO_discretization}
\end{figure}

It is demonstrated that building an IO ring array PMA is practical. To make building the magnet practical, the size of the basic building block should be constrained for safety issue. Fig.\,\ref{fig:labIO_discretization} shows an illustration. A magnet block in an IO ring array can be implemented by a row of smaller magnet columns and such a column consists of magnet cuboids, the basic magnet blocks. The size of each cuboid, $a$, $b_1$, and $h_1$ are suggested to be constrained to below 20\,mm to be handled safely by hands in a research lab. The ratios, $b_1/a$ and $b_1/h_1$, are usually constrained by the magnet manufacturer. Stacking the basic blocks to form a column is straightforward as there is no repelling force. When populated the columns around a disk housing, a locking mechanism is necessary as the columns will repel each other when they are positioned side by side. Two columns positioned right next to each other (without space) is not advised due to the large repelling force between them. If a wider magnet column is preferred, a basic block with an bigger $a$ can be considered. The ring disks (i.e., ring housing populated with magnet columns) are easy to assemble because the force toward each other is small (usually less than 10 N).

\section{Comparison between IO ring and Halbach array}
\label{sec:Halbach_IO_comp}

As introduced previously, Halbach array is a well-known PMA that has been used in dedicated portable MRI systems in recent years. A key difference between the Halbach array and an IO ring array is the field direction. A Halbach array has a transversal field direction thus the choice for RF coil is limited to a solenoid. In turn, parallel imaging with coil arrays is challenging to be implemented. An IO ring array has a longitudinal field direction which does not limit the choice of RF coils and parallel imaging can be expected when such a magnet is used to supply the main field. The second key difference is whether there is an accessible region at the central region an array. The accessible region at the center of an IO ring array allows potential inventions. 

\begin{figure}[t]
\centering
\begin{center}
\newcommand{\patchSize}{2.45cm}
\scriptsize
\setlength\tabcolsep{0.1cm}
\includegraphics[width=0.7\linewidth]{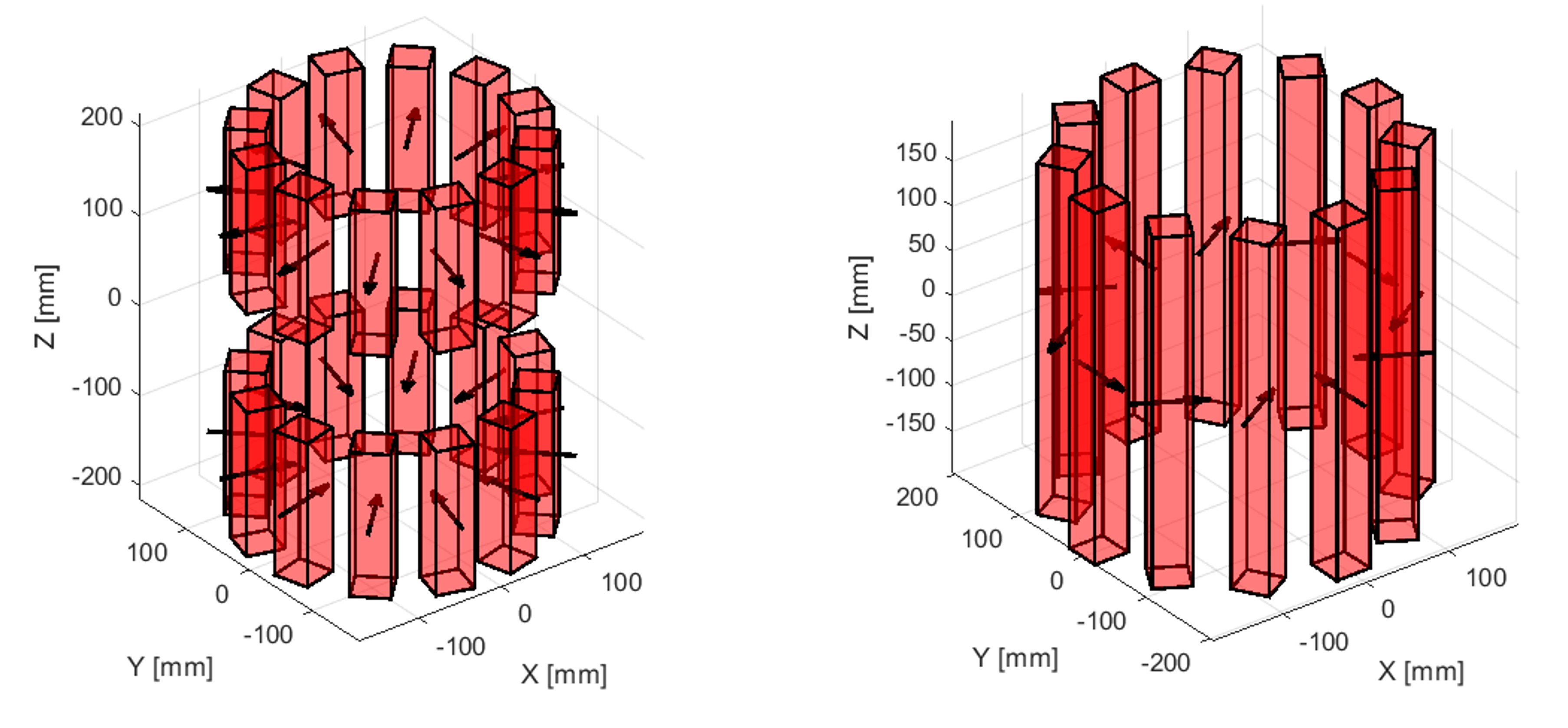}\\
(a)\hspace{0.3\linewidth}(b)\\
\includegraphics[width=0.7\linewidth]{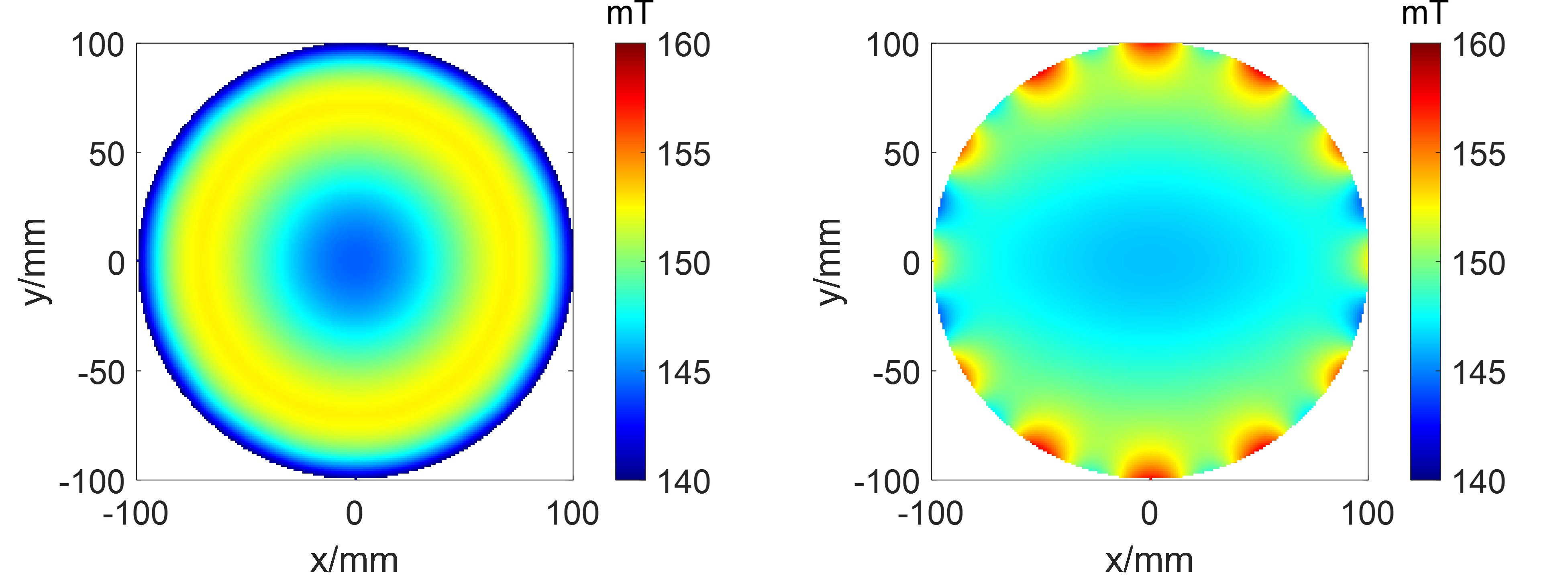}\\
(c)\hspace{0.3\linewidth}(d)\\
\end{center}
\caption{The 3D overview of (a) the IO ring array and (b) the Halbach array. (c) The $B_\text{z}$ field component from the IO ring array and (d) the $B_\text{y}$ component from the Halbach array on the $xy$-plane.}
\label{fig:field_Halbach_IO}
\end{figure}

In this section, an IO ring array PMA and a Halbach array of comparable dimensions, field strength, and field homogeneity are further compared. They are shown in Fig.\,\ref{fig:field_Halbach_IO}\,(a) and (b), respectively.
To have comparable dimensions, the two types of PMA has the following common design parameters: $n_\text{bar}=12$, $B_\text{r}=1.43$\,T, and $a=h_\text{r}=40$\,mm. The IO ring array has $h_\text{z}=160$\,mm. Both cases are optimized to have comparable field strength and field homogeneity. The IO ring arrays has variable of $R_\text{in}$ and $Z_\text{in}$ and the Halbach has variables of $R_\text{in}$ and $h_\text{z}$. They were both optimized using GA. 

The optimized parameters $R_\text{in}=120$\,mm, $Z_\text{in}=110$\,mm for the IO ring array and $R_\text{in}=146$\,mm and $h_\text{z}=390$\,mm for the Halbach array.
Fig.\,\ref{fig:field_Halbach_IO} (c)-(d) shows the optimized fields, $B_\text{z}$ of the IO ring array in Fig.\,\ref{fig:field_Halbach_IO} (c) and $B_\text{y}$ of the Halbach array, the rest of the field components are negligible. The performances of the arrays are summarized in Table\,\ref{tab:Halbach_IO}. As shown in Table\,\ref{tab:Halbach_IO}, to achieve similar field strength and homogeneity, the IO ring array uses less magnets (49.2 kg) than the Halbach array (59.9 kg). It has a higher field-strength-to-weight ratio. The downside is that the IO ring array is longer by 40\,mm and the 5-Gauss region is slightly bigger. For an IO ring array, a big 5-gauss region can be reduced by adding iron yoke between the inward and the outward magnets, which is discussed in details in Section\,\ref{subsec:with_yoke}.


\begin{table}[t]
\begin{center}
\begin{tabular}{|l|l|l|l|l|}
\hline
Case    & \begin{tabular}[c]{@{}l@{}}mean($B$)\\ {[}mT{]}\end{tabular} & \begin{tabular}[c]{@{}l@{}}Inhomogeneity\\ {[}\%{]}\end{tabular} & \begin{tabular}[c]{@{}l@{}}5-Gauss region\\ {[}mm$^3${]}\end{tabular} & Weight {[}kg{]} \\ \hline
IO ring & 149.1                                                                 & 8.7                                                              & $1280\times1280\times1600$                                            & 49.2            \\ \hline
Halbach & 149.0                                                                 & 8.9                                                              & $800\times1120\times1240$                                             & 59.9            \\ \hline
\end{tabular}
\end{center}
\caption{The average field strength, inhomogeneity, and 5-Gauss region for the comparison between the IO ring array and the Halbach array in Fig.\,\ref{fig:field_Halbach_IO}.}
\label{tab:Halbach_IO}
\end{table}



\section{Discussions}
\label{sec:discussion}

The choice of housing material for a PMA is determined by the force between the magnets. An IO ring PMA does not have stringent requirement on tensile strength of the housing material.
The magnets in either the inward or outward ring experience identical outward forces.
For a whole magnet array, as the IO ring array is axially symmetric, the force experienced by all magnets has the same magnitude and axially symmetric directions. Therefore, the force exerted on the housing structure is distributed evenly. For magnet assembly, force between two successive magnets need to be checked. In an inward or outward ring, the neighboring magnets have their polarization differed only by a small angle, which leads to big repelling force between each other. The big force can cause assembly difficulty when $n_\text{bar}$ is big and two successive magnets are close to each other. Design considerations to design one can be built in the lab without advanced tooling can be found in Section\,\ref{sec:lab_IO_construct}. If big magnet blocks are used in the design, engaging magnet assembly professionals is necessary.


The magnetic field a magnet block experience in a PMA decides the demagnetization of the magnet, and thus the stability of the magnetic field a magnet block offers in a long run. If the direction of the magnetic field exerts to a magnet differs from its own polarization by an angle $\phi$, bigger is $\phi$ stronger is demagnetization to the magnet.
For the same IO array and Halbach array as in Fig.\,\ref{fig:field_Halbach_IO}\,(a) and (b), the following calculation was performed to examine the demagnetization of magnets in the arrays. Let one magnet to be the target, and all the other magnets be the source. For each magnet, the magnetic field generated by the source at the center of the magnet was calculated and $\phi$ was calculated by comparing the direction of the calculated magnetic field to the direction of the polarization. For the IO array, due to the axial symmetry of the structure, $\phi=176.9\degree$ for every magnet. For the Halbach array, among the twelve magnets, four of them have $\phi=119.8\degree$, another four has $\phi=59.8\degree$, and a pair of $\phi=0$ and another pair of $\phi=180\degree$. It can be seen that all the magnet in the IO ring have the same and strong tendency to experience a demagnetizing field, while only a small portion of magnets in the Halbach array have this issue.
Therefore, for IO ring array, high-coercivity magnetic materials is required, such as NdFeB magnets (800-950\,kA/m\,\cite{Fuerst1993NdFeB}) and SmCo magnets (3200\,kA/m\,\cite{deCampos1998SmCo}), and magnets made of low-coercivity materials like Alnico (30-150\,kA/m\,\cite{Alnico2003}) is not suitable.

\section{Conclusion}
\label{sec:conclusion}

In this article, the properties of IO ring array are examined thoroughly via checking, 1) the relation between the design parameters $R_\text{in},Z_\text{in},h_\text{r},h_\text{z}$ and the resultant field pattern, 2) the variants and their properties. 
It is found that strong homogeneous field with high longitudinal gradients can be obtained by tuning the design parameters.
For the variants of IO ring array, it is found that the partial IO ring array can provide gradient field with high linearity. It is also found that ferromagnetic yokes can be combined with IO ring for improved homogeneity without increasing the size of the array. 
A comparison between an IO ring array and a Halbach array of comparative dimensions is presented and analyzed to show the pros and cons of these two arrays for dedicated portable MRI. 
With comparable field strength and homogeneity, IO ring array is lighter, has high field-strength-to-weight ratio besides a longitudinal field direction and accessibility in the middle of the array. 
The downside of IO ring arrays is a bigger 5-gauss zone which can be reduced by adding iron yokes between the inward and the outward magnets. Additionally, magnet blocks in an IO ring array experience stronger demagnetization than those in a Halbach array, thus magnets with high coceivity are preferred.  
Furthermore, the feasibility of building an IO ring array in a lab is introduced and demonstrated. 
The investigation strongly suggestions that an IO ring array can be a good candidate to supply the main magnetic field or the main field plus gradient fields in various dedicated MRI applications with portability.

\bibliography{sample}

\end{document}